\documentclass[aps,prd,10pt,showpacs,amsmath,showkeys,twocolumn,floatfix,amssymb, preprintnumbers, nofootinbib, superscriptaddress]{revtex4-1} 
\usepackage{epsfig,dcolumn}
\usepackage{graphicx}
\usepackage{comment} 
\usepackage[usenames]{color}
\usepackage{graphicx}
\usepackage{bm}
 \usepackage{ifpdf}
  \usepackage{floatrow}
\usepackage{makecell}
 \usepackage{caption}

\usepackage[normalem]{ulem}
\usepackage[dvipsnames]{xcolor}
\usepackage[utf8]{inputenc}
\usepackage{hyperref}
\hypersetup{ 
  pdfnewwindow=true,      
  colorlinks=true,        
  linkcolor=PineGreen,    
  citecolor=PineGreen,    
  filecolor=PineGreen,    
  urlcolor=PineGreen      
}

\newcommand{\be}{\begin{eqnarray}}
\newcommand{\ee}{\end{eqnarray}}

 \usepackage{epsfig,dcolumn}
\usepackage{graphicx}
\usepackage{comment} 
\usepackage[usenames]{color}
\usepackage{bm}
\usepackage{ifpdf}
\usepackage{floatrow}
\usepackage{makecell}
\usepackage{caption}
 
\begin{document}

\title{Visualizing resonances in finite volume}

\author{Peng~Guo}
\email{pguo@csub.edu}

\affiliation{College of Physics, Sichuan University, Chengdu, 610065, China}
\affiliation{Department of Physics and Engineering,  California State University, Bakersfield, CA 93311, USA}
\affiliation{Kavli Institute for Theoretical Physics, University of California, Santa Barbara, CA 93106, USA}

\author{Bingwei~Long}
\email{bingwei@scu.edu.cn}
\affiliation{College of Physics, Sichuan University, Chengdu, 610065, China}

\date{\today}

\begin{abstract} 
In present work, we explore and experiment an alternative approach of studying resonance properties in finite volume. By analytic continuing finite lattice size $L$ into complex plane, the oscillating behavior of finite volume Green's function is mapped into infinite volume Green's function corrected by exponentially decaying finite volume effect. The analytic continuation technique thus can be applied to study resonance properties directly in finite volume dynamical equations.  
\end{abstract}

\maketitle

\section{Introduction}\label{intro}  

The Dalitz plot is a powerful tool in particle physics to extract information from processes involving three-particle final states. For instance, the $u$- and $d$-quark mass difference can be extracted by the Dalitz analysis of  \mbox{$\eta \rightarrow 3 \pi$}   \cite{Kambor:1995yc,Anisovich:1996tx,Schneider:2010hs,Kampf:2011wr,Guo:2015zqa,Guo:2016wsi,Colangelo:2016jmc}.   Since many resonances emerge in few-hadron  systems,  the Dalitz plot also plays an important role in the study of resonance dynamics from experimental data,  {\it e.g.}   coupled-channel analysis of $\rho$ and $K^*$ resonances dynamics in \cite{Guo:2010gx,Guo:2011aa}.

On the theory side, lattice Quantum Chromodynamics (LQCD) has been one of the promising ab-initio  methods to provide understanding of few-hadron dynamics from the Standard Model. In past few years, many  progresses have been made in LQCD calculations towards understanding multi-hadron systems  \cite{Aoki:2007rd,Feng:2010es,Lang:2011mn,Aoki:2011yj,Dudek:2012gj,Dudek:2012xn,Wilson:2014cna,Wilson:2015dqa,Dudek:2016cru,Beane:2007es, Detmold:2008fn, Horz:2019rrn,Andersen:2018mau,Horz:2020zvv,Brett:2018jqw,Alexandrou:2017mpi,Fischer:2020bgv,Andersen:2017una}.
However, LQCD normally puts out discrete energy spectra of few-hadron systems, because of the finite volume inherent to the method, rather than reaction amplitudes which are needed to generate the Dalitz plot. 
Lattice QCD calculations are usually performed with spatial periodic boundary condition.
In the two-body sector, connections between infinite-volume reaction amplitudes and energy levels in a cubic box under periodic boundary condition can be constructed in a compact and elegant  equation, normally referred as the  L\"uscher  formula \cite{Luscher:1990ux}, and it has since been    extended to cases including moving frames and coupled channels  \cite{Rummukainen:1995vs,Christ:2005gi,Bernard:2008ax,He:2005ey,Lage:2009zv,Doring:2011vk,Briceno:2012yi,Hansen:2012tf,Guo:2012hv,Guo:2013vsa,Agadjanov:2016mao}.

Various approaches on finite-volume three-particle dynamics exist \cite{Kreuzer:2008bi,Kreuzer:2009jp,Kreuzer:2012sr,Polejaeva:2012ut,Briceno:2012rv,Hansen:2014eka,Hansen:2015zga,Hansen:2016fzj,Briceno:2017tce,Sharpe:2017jej,Hammer:2017uqm,Hammer:2017kms,Meissner:2014dea,Mai:2017bge,Mai:2018djl, Doring:2018xxx, Romero-Lopez:2018rcb,Guo:2016fgl,Guo:2017ism,Guo:2017crd,Guo:2018xbv,Blanton:2019igq,Romero-Lopez:2019qrt,Blanton:2019vdk,Mai:2019fba,Guo:2018ibd,Guo:2019hih,Guo:2019ogp,Guo:2020wbl,Guo:2020kph}, such as   the relativistic all-orders perturbation theory  \cite{Hansen:2014eka,Hansen:2015zga,Hansen:2016fzj,Briceno:2017tce,Sharpe:2017jej}, effective theory based approach \cite{Kreuzer:2008bi,Kreuzer:2009jp,Kreuzer:2012sr,Hammer:2017uqm,Hammer:2017kms,Klos:2018sen}, and Faddeev type equation based variational approach  \cite{Guo:2018ibd,Guo:2019hih,Guo:2019ogp,Guo:2020wbl,Guo:2020kph}.  As pointed out in  Ref.~\cite{Guo:2020iep}, though the quantization conditions  are formulated in different ways depending on a specific approach, all approaches  share the same strategy and similar features. The infinite-volume reaction amplitudes are in fact not directly extracted from lattice data. Only subprocess interactions or associated subprocess amplitudes are obtained from quantization condition, and total infinite volume reaction amplitudes have to be computed in a separate step.  Two-step procedures seems like a compromised solution, and one may hope to have ultimate formalism that grant user direct access to infinite-volume reaction amplitudes from the LQCD energy levels, independently of inter-hadron interaction models. But deriving such relations beyond two-body systems  poses great challenges. With  two-step procedures,  the  finite-volume and infinite-volume physics  can   be dealt separately,  eventually linked by inter-hadron interactions at one's proposal.  Therefore, the quantization condition is free of infinite-volume reaction amplitudes and it  is more straightforward to implement in practical data analysis of LQCD results.  The model dependence of the inter-hadron potential can be assessed by how well it fits to the LQCD energy levels. However,scattering observables, such as  the Dalitz plot, must be computed in a separate step.

In present work,    we aim to explore and experiment an alternative approach  by fully taking   advantage of Faddeev type integral equations in finite volume. By analytic continuation of box size $L$ into complex plane, the mapping relation of finite volume Green's function in different energy domains  can be established as the consequence of global spatial symmetry. Therefore, in finite volume, $L$ may be used as a tunable parameter to turn an oscillating finite volume Green's function into the infinite volume Green's function with some exponentially decaying finite volume corrections. Hence,  the resonance show up as a peak as well in finite volume scattering amplitudes even for small $L$ values.

The paper is organized as follows. The analytic continuation technique is explained and  demonstrated in Sec.\ref{fvTanalycont}  with the  one-dimension case. 
We summarize the findings in Sec.~\ref{summary}.

\section{Analytic continuation of finite volume amplitudes}\label{fvTanalycont}

The finite-volume Green's function is qualitatively different from its infinite-volume counterpart in that one has poles in the complex energy plane corresponding to discrete levels, and the other has branch cuts corresponding to continuous spectrum. So a very large cubic box must be used if one would like to approximate scattering states in finite volume. However, as we will show in this section, if $L$ is analytically continued to its complex plane, finite-volume amplitudes can resemble quite well actual reaction amplitudes even for relatively small values of $L$.    To keep the technical discussion at a manageable level,    our presentation in follows will be only limited to non-relativistic few-body system in one-dimensional space. The physical application of such a technique, such as the Dalitz plot of $\eta, \eta' \rightarrow 3\pi$, will be  presented in a separate work.

\subsection{Finite volume amplitude and L\"uscher  formula }

 Let us start  the discussion with two-body interactions in finite volume. One of major  tasks of investigating finite-volume dynamics is to look for the discrete eigen-energies of few-body systems in a periodic box. 
These energy eigen-states are stationary and are described by a homogeneous Lippmann-Schwinger (LS) equation in the two-body case:
\begin{equation}
\phi_L (E)=  G_L (E)V \phi_L (E), 
\end{equation}
where $G_L$ and $\phi_L$ stand for the finite volume Green's function and wave function respectively, and $V$ denotes the interaction potential of particles.   Equivalently,  the homogeneous LS equation can be rewritten as
\begin{equation}
t_L (E)= V G_L (E) t_L (E), 
\end{equation}
where   $$t_L (E) = - V \phi_L (E).$$ 
The  energies of stationary states are   those letting the following determinant vanish,
\begin{equation}
\det \left [ I- V G_L (E) \right ]=0, \ \ \ \  E \in \{ E_1, \cdots , E_n , \cdots \}. \label{detqc2b}
\end{equation}

It  is useful to introduce  an operator $\tau_L (E)$ that satisfies  the inhomogeneous LS equation,
\begin{equation}
\tau_L (E)= -  V+V G_L (E) \tau_L (E) \, . \label{tau2beq}
\end{equation}   
 The solution of Eq.(\ref{tau2beq}) is symbolically given  by 
\begin{equation}
\tau_L (E) = - \frac{1}{\frac{1}{V} - G_L(E)},
\end{equation} 
which  will be used in  the three-body finite-volume LS equations. The poles of  $\tau_L (E) $ amplitude also   yield  eigenenergy solutions of stationary state  of the few-body finite volume system, which is consistent with  the quantization condition given by Eq.(\ref{detqc2b}). 

 The matrix element of $\tau_L (E)$ between two plane waves defines the   finite-volume transition amplitude, which could be on-shell or off-shell.
Using 1D as  the example,  the off-shell finite-volume amplitude  in plane wave basis is given by
\begin{equation}
\tau_L (k; E; k') = \langle k | \tau_L (E) | k'  \rangle \, ,
\end{equation} 
where 
\begin{equation}
(k, k' ) \in \frac{2\pi n}{L}, \; \text{and} \ \ n\in \mathcal{Z}.
\end{equation}
 Eq.(\ref{tau2beq}) thus yields
 \begin{align}
 & \tau_L (k; E;k')   = - \widetilde{V} (k-k')  \nonumber \\
 &+ \sum_p \widetilde{V} (k- p) \widetilde{G}_L (p; E)  \tau_L (p; E; k') , \label{tau2boffeq}
 \end{align}
 where  the two-body finite-volume Green's function in  the CM frame is given by
 \begin{equation}
\widetilde{G}_L (p ; E)  = \frac{1}{L} \frac{1}{ m E -   p^2 }.
\end{equation}
 Equation~\eqref{tau2boffeq} resembles the  LS equation in infinite volume for scattering states, hence discrete $ (k' , k)\neq \sqrt{mE}$ may be interpreted as off-shell  incoming    and outgoing momenta respectively.

 Considering short-range interaction approximation, $\widetilde{V} (k-k')  \simeq \widetilde{V} (0)= V $,   the solution to Eq.(\ref{tau2boffeq}) is thus dominated primarily by diagonal terms of off-shell amplitudes,  $ \tau_L (k; E;  k)   \simeq  \tau_L ( E)  $,   which is normally also referred  as on-shell approximation,   see Refs. \cite{Bernard:2008ax,Doring:2013glu}. Hence, Eq.(\ref{tau2boffeq}) is reduced to a algebra equation, and the solution is given by
  \begin{equation}
     \tau_L ( E)  = - \frac{1}{\frac{1}{V} - G_L(0,E)}   , \label{tau2boneq}
  \end{equation}
  where 
   \begin{equation}
  G_L (0, E) = \frac{1}{L} \sum_{p = \frac{2\pi n}{L}, n\in \mathbb{Z}}   \frac{1}{ m E -   p^2 } = \frac{ \cot \frac{ \sqrt{m E  }}{2} L }{2 \sqrt{m E  }} .
 \end{equation}
 The potential term  $\frac{1}{V} $ is related to scattering phase-shift and infinite volume Green's function,
    \begin{equation}
  \frac{1}{V} = - \frac{1}{2 \sqrt{mE}}  \left [ \cot \delta(E) - i \right ] + G_\infty(0,E)  , 
  \end{equation}  
  where
   \begin{equation}
  G_\infty (0, E) =  \int \frac{d p}{2\pi}  \frac{1}{ m E -   p^2   } = -\frac{ i }{2 \sqrt{m E}}. \label{Ginf1D}
   \end{equation}
  Thus, Eq.(\ref{tau2boneq}) can be rewritten in the form that is associated to L\"uscher  formula, 
   \begin{equation}
     \tau_L ( E)  =  \frac{1}{2 \sqrt{mE}  }  \frac{1}{   \cot \delta(E)  - \mathcal{M}_L (E)}   ,
  \end{equation}
   where   $ \mathcal{M}_L (E)$ is L\"uscher's zeta function,
  \begin{equation}
 \mathcal{M}_L (E) = i+    2 \sqrt{mE} \left[ G_\infty(0,E) - G_L(0,E) \right ].
  \end{equation} 
    The pole of  $ \tau_L ( E)  $ yields  L\"uscher  formula 
   \begin{equation}
   \cot \delta(E)  - \mathcal{M}_L (E)=0.
   \end{equation}

\begin{figure}
\begin{center}
\includegraphics[width=0.88\textwidth]{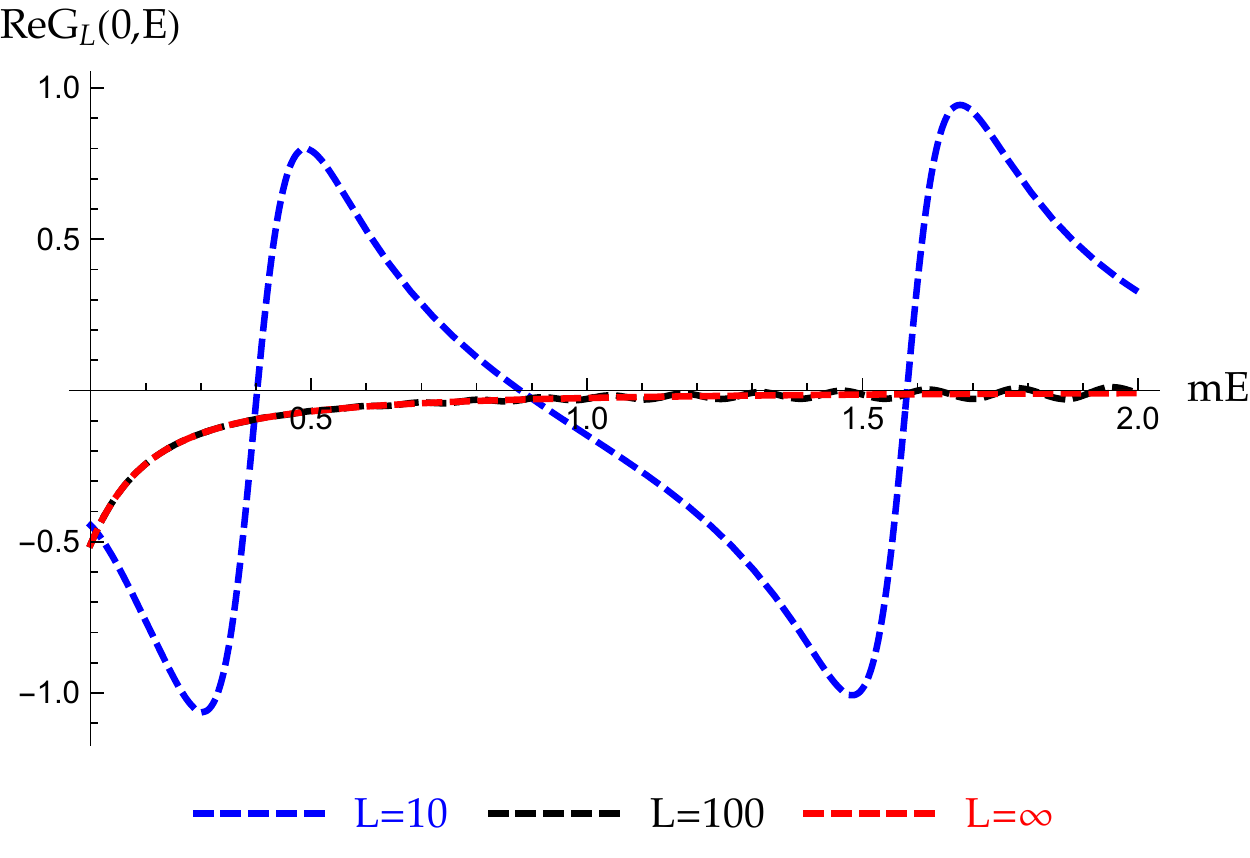}
\includegraphics[width=0.88\textwidth]{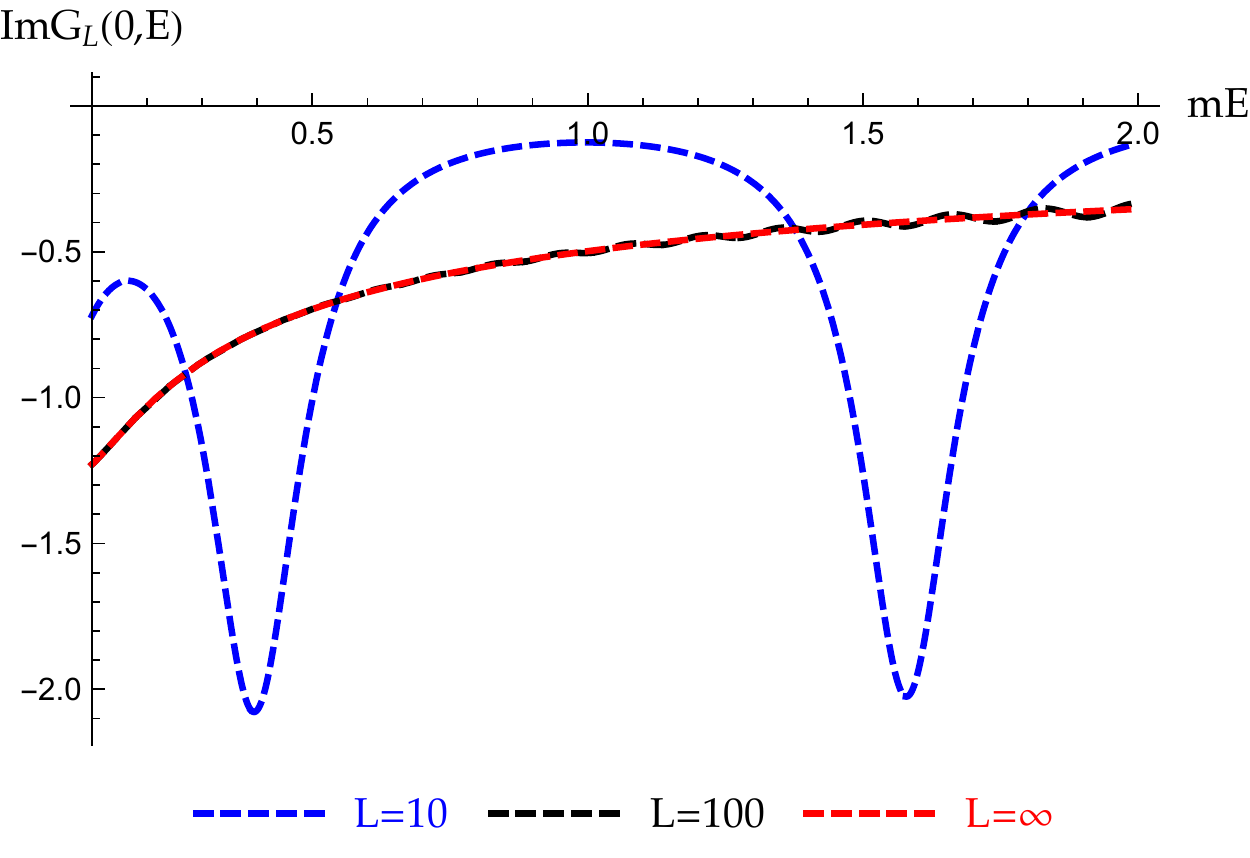}
\caption{ The comparison of    $ G_L (0, E) $ and $G_\infty (0, E)$ with complex argument $m E + i \epsilon$, where $\epsilon=0.1$ and  $L=10 $ (blue), $100$ (black) and $\infty$ (red). }\label{G2bfvinfvplot}
\end{center}
\end{figure}

    Although the expression of $\tau_L (E) $ in Eq.(\ref{tau2boneq})   has similar structure  as its infinite volume counterpart,  the on-shell two-body scattering amplitude,
 \begin{equation}
 \tau_\infty (E) =  - \frac{1}{ \frac{1}{ V}   -  G_\infty (0, E) } = \frac{1}{2 \sqrt{mE}  }  \frac{1}{   \cot \delta(E)  - i} , 
 \end{equation} 
 $ \tau_L (E) $ and $ \tau_\infty (E) $ behave significantly different due to superficially divergent analytic appearance of finite volume and infinite volume Green's functions. In finite volume, $  G_L (0, E) $
 is a periodic oscillating real function, compared with infinite volume counterpart $G_\infty (0, E)$
which is purely imaginary. The difference between $ G_L (0, E) $ and $G_\infty (0, E)$ may be understood from analytic properties of Green's functions, in infinite volume, Green's function is determined by the branch cut  lying on positive real axis,
 \begin{equation}
  G_\infty (0, E) =   \frac{1}{2\pi} \int_0^\infty d s'   \frac{\sqrt{\frac{1}{s'}}}{ m E -   s'  }  .
   \end{equation}
For the values of $mE $ taking slightly above real axis by $m E + i \epsilon$,   principle part of above interaction  vanishes, only  absorptive part survives that yields imaginary part $-\frac{ i }{2 \sqrt{m E}}$.  However, in finite volume, the branch cut dissolve into discrete poles lying on real axis,  therefore, for the values of $mE$ not overlapping with pole positions, only principle part contributes. It is also interesting to see that for finite $i \epsilon$,  both $ G_L (0, E) $ and $G_\infty (0, E)$  becomes complex, and sharp oscillating behavior of $ G_L (0, E) $ is smoothed out and start matching with  $G_\infty (0, E)$ when $L \gg 1/\sqrt{\epsilon}$, see  Fig.~\ref{G2bfvinfvplot} as a example. In other words, as $\epsilon \rightarrow 0$, $ \tau_L (E) $ and $ \tau_\infty (E) $ indeed behaves significantly different for finite $L$, and $ \tau_L (E) \rightarrow \tau_\infty (E) $ only when $L \rightarrow \infty$. Therefore,  $ \tau_L (E) $ for finite $L$  normally are not considered as a useful tool for the identification of a sharp resonance    that on the contrary appear as a peak in $ \tau_\infty (E) $.

Next, we will explain how  analytic continuation technique may   allow one to have finite volume amplitude that resemble the behavior of infinite volume amplitude for even finite $L$   and    real  $ mE $  values with $\epsilon \rightarrow 0$. It turns out that analytic continuation technique is the direct consequence of global symmetry of Green's function in complex space.

\subsection{Global symmetry and analytic continuation of Green's function}
Using again 1D non-relativistic few-body system as example, in infinite volume, two-body Green's function satisfies
\begin{equation}
\left ( m E + \frac{d^2}{d x^2} \right ) G_\infty (x, E) = \delta(x), \label{infGreen1D}
\end{equation}
where the physical value of position $x$ is defined on real axis. Now, let's extend $x$ to complex plane by multiplying a phase factor $e^{i \theta}$, 
$$x \rightarrow x e^{i \theta},$$ 
where $\theta \in [0, \pi]$, we remark that the method of extension of  $x$ to complex plane is also known as complex scaling method in nuclear and atomic physics, see Refs.~\cite{HO19831,MOISEYEV1998212,moiseyev_2011}. In complex space, Eq.(\ref{infGreen1D}) yields
\begin{equation}
\left ( m E e^{i 2 \theta} + \frac{d^2}{d x^2} \right ) \left [ e^{-i \theta} G_\infty (x e^{i \theta}, E) \right ] = \delta(x). \label{infGreen1Dcomplex}
\end{equation}
One can conclude that Green's function equation is invariant under global rotation of space in complex plane, and $ e^{-i \theta} G_\infty (x e^{i \theta}, E)$ is related to the solution of Eq.(\ref{infGreen1D}) on real axis with  eigenenergy of $m E e^{i 2 \theta}$, hence
\begin{equation}
G_\infty (x e^{i \theta}, E) = e^{i  \theta}G_\infty (x, E e^{i 2 \theta}) .
\end{equation}
For $\theta= \frac{\pi}{2}$, thus
\begin{equation}
G_\infty (0, E) = i G_\infty (0, - E  ) , \label{GinfEmap}
\end{equation}
this is indeed consistent with analytic expression in Eq.(\ref{Ginf1D}). That is to say that the Green's function in physical region $(E>0)$ is mirrored to unphysical region ($E<0$) by Eq.(\ref{GinfEmap}) as the consequence of global spatial symmetry. In infinite volume,   $E$ is only tunable parameter in $G_\infty (0, E) $ that can be used to cross from physical region to unphysical region or vice versa.

The finite volume two-body Green's function is given by
\begin{equation}
\left ( m E + \frac{d^2}{d x^2} \right ) G_L (x, E) =\sum_{n \in \mathcal{Z}} \delta(x+ n L). \label{fvGreen1D}
\end{equation}
Similarly, we want to extend finite system to complex   plane by a global spatial rotation,  $x \rightarrow x e^{i \theta}$ and also  $L \rightarrow L e^{i \theta}$, thus we find
\begin{equation}
\left ( m E e^{i 2 \theta} + \frac{d^2}{d x^2} \right ) \left [ e^{-i \theta} G_{L e^{i \theta}  } (x e^{i \theta}, E) \right ] =\sum_{n \in \mathcal{Z}} \delta(x+ n L). \label{fvGreen1Dcomplex}
\end{equation}
 Eq.(\ref{fvGreen1Dcomplex}) yields  a useful relation
\begin{equation}
G_{L e^{i \theta}  } (0, E)  = e^{i  \theta}G_L (0, E e^{i 2 \theta}) . \label{GfvEmap}
\end{equation}
A key observation is that because of extra tunable parameter $L$ in finite volume, now, Eq.(\ref{GfvEmap}) allow one to navigate freely between physical and unphysical regions of finite volume Green's function with a fixed  value of $E$.  

  \begin{figure}
\begin{center}
\includegraphics[width=0.9\textwidth]{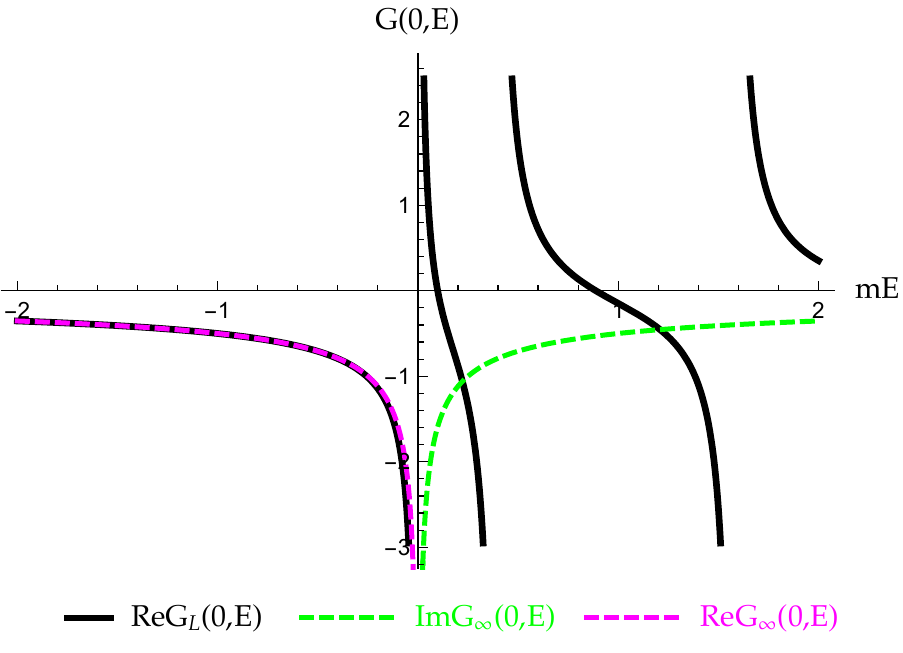}
\includegraphics[width=0.9\textwidth]{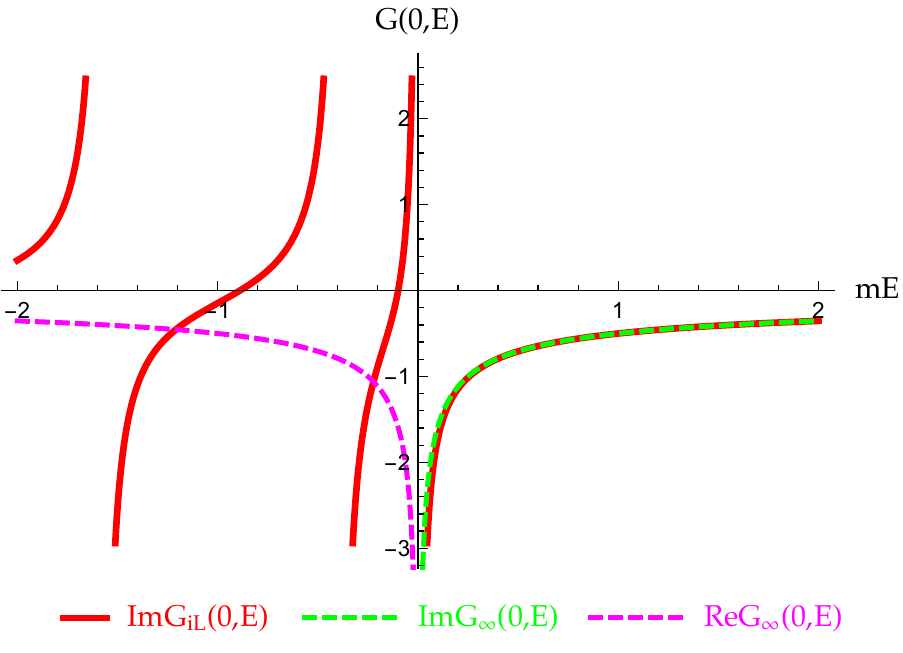}
\caption{  Plot of $Re G_L(0,E)$ (black solid) and $Im G_{i L} (0,E)$ (red solid) with $L=10$ vs. $Im G_\infty(0,E)$ (dashed green) and $Re G_\infty(0,E)$ (dashed magenta).}\label{GL1Dplot}
\end{center}
\end{figure}

It is also worth noticing that though $G_L(0, E)$ and $G_\infty(0, E)$ are  significant different for   $E>0$ with finite $L$,  below physical threshold ($E<0$), 
\begin{equation}
G_L(0, E < 0 ) = -  \frac{\coth \frac{\sqrt{|mE|} L}{2}}{2\sqrt{|mE|}}
\end{equation}
 quickly approaches its infinite volume counterpart  $G_\infty(0, E<0) = - \frac{1}{2\sqrt{|mE|}}$ as increasing $L$, due to the fact that 
 \begin{equation}
 \coth \frac{ \sqrt{m E  }}{2} L = 1 +2 e^{ - \sqrt{m E  }  L} + 2 e^{ -2  \sqrt{m E  }  L} + \cdots \stackrel{L \rightarrow \infty}{\rightarrow}   1.
 \end{equation}
Now, using relation in  Eq.(\ref{GfvEmap}), simply rotating $L \rightarrow i L $, we find
 \begin{equation}
  G_{i L} (0, E) =  - i \frac{ \coth \frac{ \sqrt{m E  }}{2} L }{2 \sqrt{m E  }}  = G_{\infty} (0, E) \coth \frac{ \sqrt{m E  }}{2} L  .
 \end{equation}
Therefore, in physical region $E>0$,    $G_{i L} (0, E) $ behaves just as $G_\infty(0, E)  $ with exponential decaying corrections due to finite volume effect, see Fig.\ref{GL1Dplot}.

\subsection{Resonance in  $ i L$ space}

 With continuation $L \to iL$, the finite-volume on-shell amplitude
\begin{equation}
\tau_{i L} (E) = \frac{1}{2 \sqrt{mE}  }  \frac{1}{   \cot \delta(E)  - \mathcal{M}_{iL} (E)}  
\end{equation}
 approximates very well $\tau_\infty(E)$  in the physical region of $E$,  differing only by powers of $e^{ - \sqrt{m E  }  L}$:
   \begin{equation}
 \mathcal{M}_{iL} (E) =  i \left ( 1 +2 e^{ - \sqrt{m E  }  L} + 2 e^{ -2  \sqrt{m E  }  L} + \cdots \right )  \stackrel{L \rightarrow \infty}{\rightarrow}   i .
  \end{equation} 
 We will illustrate how this helps identify resonances as a peak of $\tau_{i L} (E)$, just as in the infinite-volume case.

\begin{figure}
\begin{center}
\includegraphics[width=0.88\textwidth]{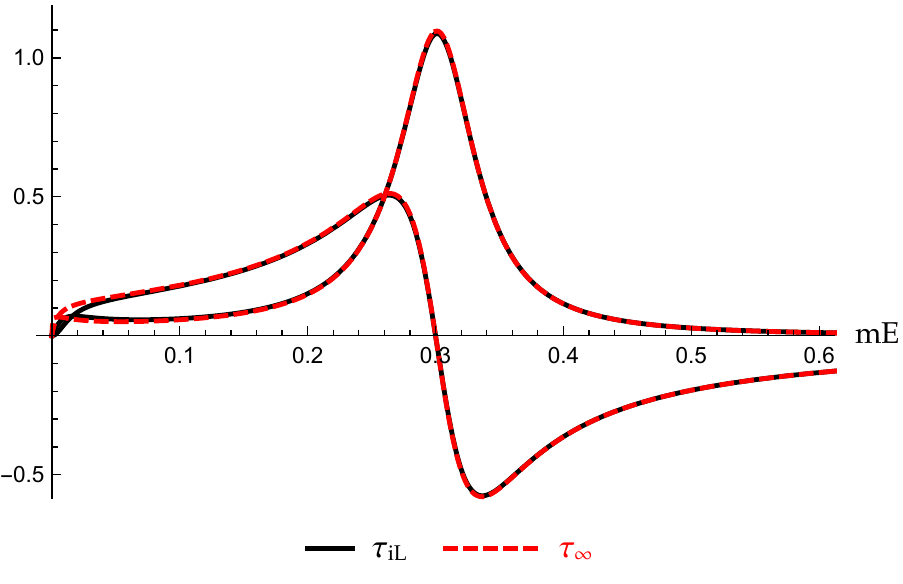}
\caption{ The comparison of    $ \tau^{(res, 0)}_{ i L}  (0, E) $ (solid black) and $ \tau^{(res, 0)}_{ \infty}  (0, E)$ (dashed red) with    $L=10 $, $mV_0 =0$, $g_\rho = 0.04$ and $m m_\rho = 0.3$. }\label{tau2biLplot}
\end{center}
\end{figure}

 We propose a resonance model by replacing $V$ by
$$
m V_0 +  \frac{g_\rho}{m (E -  m_\rho)} \, ,
$$
where  $g_\rho$ and $m_\rho)$  are the coupling constant and mass of  the resonance respectively, and $mV_0$  can be used to parameterize the background contribution.  The infinite- volume scattering amplitude is thus given  by
 \begin{equation}
    \tau^{(res)}_{ \infty} ( E)  =  - \frac{1}{ \frac{1}{m V_0 +  \frac{g_\rho}{m (E -  m_\rho )} } +   \frac{  i }{2 \sqrt{m E  }}    } ,
  \end{equation}
 to be compared with  the analytically continued finite-volume amplitude:
 \begin{equation}
    \tau^{(res)}_{ i L} ( E)  =  - \frac{1}{ \frac{1}{m V_0 +  \frac{g_\rho}{m ( E -    m_\rho )} } +    \frac{ i }{2 \sqrt{m E  }}  \coth \frac{ \sqrt{m E  }}{2} L   } . 
  \end{equation}
The comparison of $  \tau^{(res)}_{ \infty} ( E) $ and $  \tau^{(res)}_{ i L} ( E)  $ for a resonance model  is plotted in Fig.~\ref{tau2biLplot}.

\begin{figure}
\begin{center}
\includegraphics[width=0.88\textwidth]{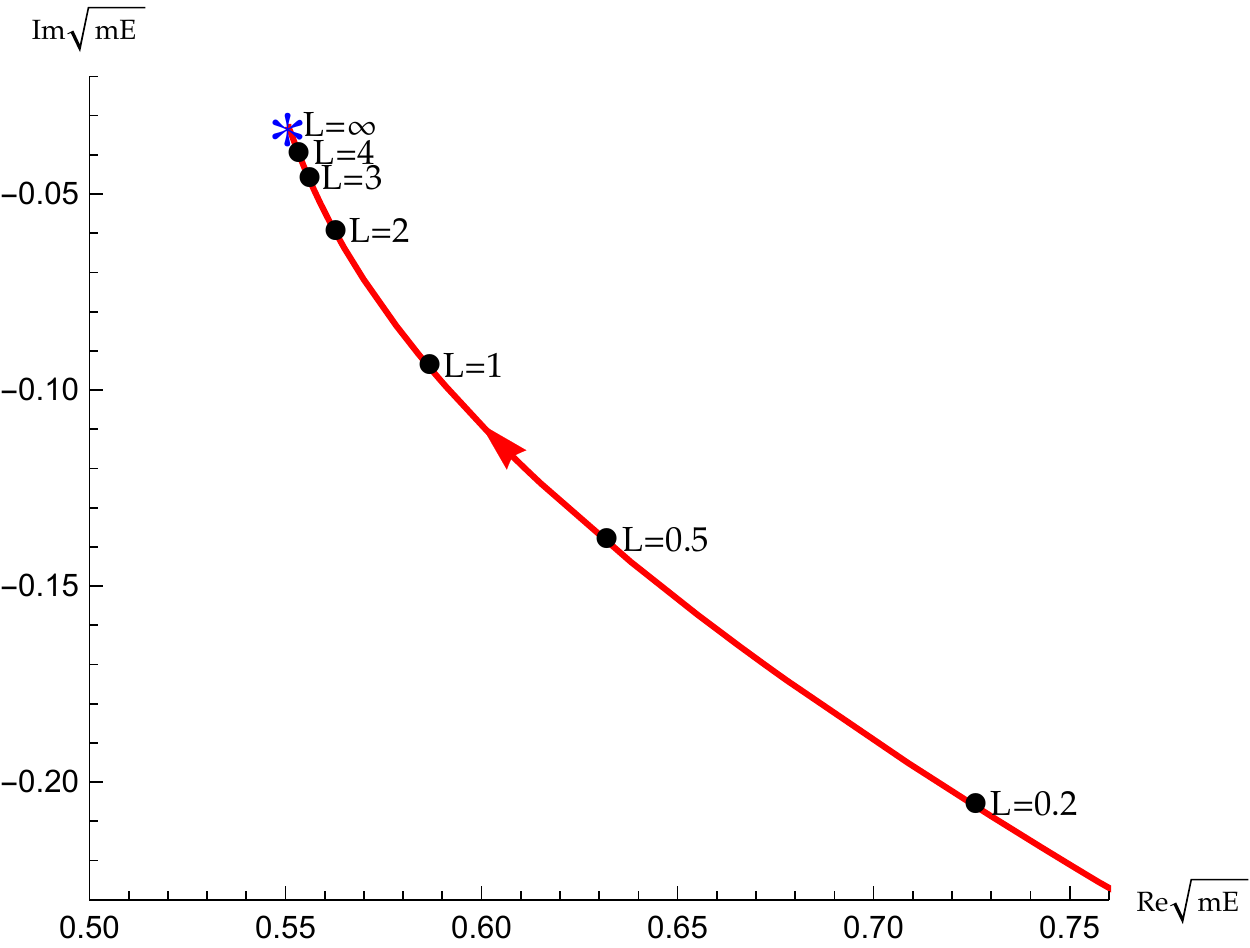}
\caption{ The  trajectory of resonance pole position as the function of $L$ with   parameters of resonance model: $mV_0 =0$, $g_\rho = 0.04$ and $m m_\rho = 0.3$.  Arrow indicates increasing $L$ direction.}\label{poleplot}
\end{center}
\end{figure}

In infinite volume,  the resonance pole position is   given by
\begin{equation}
\frac{1}{m V_0 +  \frac{g_\rho}{m (E -  m_\rho )} } +   \frac{  i }{2 \sqrt{m E  }}   =0,
\end{equation}
 and in finite volume, the pole position
  of $\tau^{(res, 0)}_{ i L} ( E) $  is shifted:  
\begin{equation}
\frac{1}{m V_0 +  \frac{g_\rho}{m ( E -    m_\rho )} } + i   \frac{ \coth \frac{ \sqrt{m E  }}{2} L }{2 \sqrt{m E  }} =0.
\end{equation}
Figure~\ref{poleplot} depicts the the resonance pole of $\tau^{(res, 0)}_{ i L} ( E) $ for various values of $L$ and the pole of $\tau^{(res)}_{\infty} (E)$. It shows clearly how rapidly the finite-volume pole approaches its infinite-volume limit.

\subsection{Inhomogeneous three-particle  Faddeev type equation in finite volume}\label{3bFaddeev}
The idea presented in previous sections can now be applied into  three-body systems. In finite volume, the stationary solutions of three-body system may be described by  an homogeneous Schr\"odinger equation,
\begin{equation}
\Phi_L = G_L (E) (V_{12} + V_{23} +V_{31}) \Phi_L, \label{PhiLeq}
\end{equation}
where the subscript of $V_{\alpha \beta } $ is used to describe the interaction among different pairs.  The Eq.(\ref{PhiLeq}) can be converted into homogeneous Faddeev type coupled dynamical equations, see \cite{Guo:2019hih,Guo:2019ogp,Guo:2020wbl,Guo:2020kph}.  In   a simple case with three identical particles interacting through only pair-wise interaction,   only one dynamical equation is required, symbolically, it is given by
\begin{equation}
T_L (E) = \frac{2}{\frac{1}{V}- G_L (E)} G_L (E)  T_L (E), \label{TLeq}
\end{equation}
where $T_L(E) = -V \Phi_L(E)$ stands for finite volume Faddeev three-body amplitude. Homogeneous Eq.(\ref{TLeq}) or Eq.(\ref{PhiLeq}) thus  yield quantization condition
\begin{equation}
\det \left [ I - \frac{2}{\frac{1}{V}- G_L (E)} G_L  (E)\right ] =0, \label{3bqcpairwise}
\end{equation}
 which determines  discrete eigenenergy of stationary solutions that satisfies periodic boundary condition in finite volume.   In addition to establishing quantization condition and obtaining eigenenergies,  the finite volume wave function may also be computed   from dynamical equations, Eq.(\ref{TLeq}) and Eq.(\ref{PhiLeq}), the technical details of extracting few-body finite volume wave function is presented in Appendix \ref{wavanlycont}.

To compare with infinite volume scattering amplitudes, let's introduce three-body  operator $\mathcal{T}_L (E) $ that satisfies inhomogeneous three-body  equation,
\begin{equation}
\mathcal{T}_L (E)  = - \frac{1}{\frac{1}{V} - G_L (E)} + \frac{2}{\frac{1}{V} - G_L (E)}  G_L (E) \mathcal{T}_L (E) , \label{Topoffsheeq}
\end{equation} 
the poles of  $\mathcal{T}_L (E) $ correspond to the stationary solutions that are also described by   homogeneous equation,  Eq.(\ref{TLeq}). The off-shell finite volume amplitude in plane wave basis that resembles scattering amplitude in infinite volume may be introduced by
\begin{equation}
 \mathcal{T}_L (k_1, k_2; E; k'_1, k'_2)  =  \langle k_1, k_2 |  \mathcal{T}_L (E)   | k'_1 k'_2 \rangle_L,
\end{equation}
where $(k_1, k_2)$ and $( k'_1,k'_2) \in \frac{2\pi}{L} n, n \in \mathbb{Z}$ represent outgoing and incoming two independent momenta of   particles. While considering only pair-wise contact interaction, $ \mathcal{T}_L (k_1, k_2; E; k'_1, k'_2) $  depends only on a single outgoing momentum, {\it e.g.} in $(13)$ channel,
\begin{align}
&  \mathcal{T}_L (k_1, k_2; E; k'_1, k'_2)   = - \int_L d r_{13} d r_{23} e^{-i k_1 r_{13}}  e^{- i k_2 r_{23}} \nonumber \\
 & \times m V_0 \delta(r_{13}) \phi_L(r_{13}, r_{23}; k'_1, k'_2) =   \mathcal{T}_L (  k_2; E; k'_1, k'_2).
\end{align}
The off-shell inhomogeneous three-body LS equation for pair-wise contact interaction is thus given in a compact form,
\begin{align}
& \mathcal{T}_L (k; E; k'_1, k'_2) = \tau_L^{(k)} (E)  \langle k | k'_1 k'_2 \rangle_L  \nonumber \\
& -2 \tau_L^{(k)} (E)   \sum_{p= \frac{2\pi n}{L} , n \in \mathbb{Z}}  L \widetilde{G}_L (k,p; E)  \mathcal{T}_L (p;E; k'_1,k'_2), \label{TLoffsheq1D}
\end{align}
where 
\begin{equation}
 \langle k | k'_1 k'_2 \rangle_L = L \left [ \delta_{k, k'_2} + \delta_{k, k'_1} + \delta_{k, k'_3} \right ].
\end{equation}
Three-body finite volume Green's function is given by
\begin{equation}
\widetilde{G}_L (p_1,p_2; E)  = \frac{2}{L^2} \frac{1}{2 m E - \sum_{i=1}^3 p_i^2 },
\end{equation}
and two-body amplitude $ \tau^{(k)}_{L} (E)  $ in a moving frame  is defined by
\begin{align}
& \tau^{(k)}_L (E)   =- \frac{1}{ \frac{1}{ m V_0} -    \sum_{p}  L \widetilde{G}_L (k,p ; E)   } \nonumber \\
& =  - \frac{1}{ \frac{1}{ m V_0} -  \frac{ \cot \frac{ \sqrt{m E - \frac{3}{4} k^2} - \frac{k}{2}}{2} L + \cot \frac{ \sqrt{m E - \frac{3}{4} k^2} + \frac{k}{2}}{2} L  }{4 \sqrt{m E - \frac{3}{4} k^2}}    } .
\end{align}
The   $k$ and $\sqrt{m E - \frac{3}{4} k^2}$  in $\tau^{(k)}_L (E) $ represent the total and the relative momenta of two-particle pair respectively.

Instead of solving off-shell LS equation, Eq.(\ref{TLoffsheq1D}), it may be more convenient to introduce a half-off-shell amplitude that only depends on outgoing momenta of particles by sum over all the initial momenta $(k'_1,k'_2)$,
\begin{equation}
\mathcal{ T}_L (k; E )  = \frac{1}{ \sum_{k'_1, k'_2}  \langle k | k'_1 k'_2 \rangle_L  }  \sum_{k'_1, k'_2} \mathcal{ T}_L (k; E; k'_1, k'_2) ,
\end{equation}
thus,  Eq.(\ref{TLoffsheq1D}) is converted into
\begin{align}
& \mathcal{T}_L (k; E ) = \tau_L^{(k)} (E)     \nonumber \\
& - 2 \tau_L^{(k)} (E)   \sum_{p = \frac{2\pi n}{L} , n \in \mathbb{Z}}  L \widetilde{G}_L (k,p; E)  \mathcal{T}_L (p;E ). \label{TLhalfoffsheq1D}
\end{align}
$\mathcal{ T}_L (k; E ) $ amplitude may be used to describe the decay process.  Eq.(\ref{TLhalfoffsheq1D}) thus resemble   isobar model in dispersion approach \cite{Kambor:1995yc,Schneider:2010hs,Guo:2015zqa,Guo:2016wsi,Guo:2014vya}, where $\tau_L^{(k)} (E)  $ may be interpreted as naive isobar pair term, and second term  in Eq.(\ref{TLhalfoffsheq1D}) thus corresponds to the isobar corrections due to rescattering effect among different isobar pairs.  Just as two-body finite volume amplitude, for finite value of $L$ and real $E$, $\mathcal{ T}_L (k; E )  $ is real and oscillating function that may not be the suitable form for the task of identification of resonance.

\subsection{Analytic continuation of three-particle Faddeev type equation in finite volume and  solutions in $iL$-space}

Inhomogeneous three-body dynamical equation, Eq.(\ref{TLhalfoffsheq1D}), can be  analytic continued into $i L$ space as well, 
\begin{align}
& \mathcal{T}_{iL} (k; E ) = \tau_{iL}^{(k)} (E)   \nonumber \\
&   -2 \tau_{iL}^{(k)} (E)   \sum_{p = \frac{2\pi n}{i L}, n \in \mathbb{Z}}   (i L ) \widetilde{G}_{iL} (k,p; E)  \mathcal{T}_{i L} (p;E ), \label{TiLeq1D}
\end{align}
where
\begin{equation}
\widetilde{G}_{iL} (p_1,p_2; E)  = \frac{2}{(i L)^2} \frac{1}{2 m E - \sum_{i=1}^3 p_i^2 }, 
\end{equation}
and
\begin{equation}
 \tau^{(k)}_{iL} (E)     =  - \frac{1}{ \frac{1}{ m V_0} + i  \frac{ \coth \frac{ \sqrt{m E - \frac{3}{4} k^2} - \frac{k}{2}}{2} L + \coth \frac{ \sqrt{m E - \frac{3}{4} k^2} + \frac{k}{2}}{2} L  }{4 \sqrt{m E - \frac{3}{4} k^2}}    } .
\end{equation}
Now, it will be illustrated in follows that the solution of Eq.(\ref{TiLeq1D}), $\mathcal{T}_{iL} (k; E ) $,  will  behave similar to its counterpart in infinite volume, $\mathcal{T}_{\infty} (k; E )$, which is given by
\begin{align}
& \mathcal{T}_{\infty} (k; E ) = \tau_{\infty}^{(k)} (E)   \nonumber \\
&   -2 \tau_{\infty}^{(k)} (E)  \int  d p  (2\pi) \widetilde{G}_{\infty} (k,p; E)  \mathcal{T}_{\infty} (p;E ), \label{Tinfveq1D}
\end{align}
where
\begin{equation}
\widetilde{G}_{\infty} (p_1,p_2; E)  = \frac{2}{(2\pi)^2} \frac{1}{2 m E - \sum_{i=1}^3 p_i^2 }, 
\end{equation}
and
\begin{equation}
 \tau^{(k)}_{\infty} (E)     =  - \frac{1}{ \frac{1}{ m V_0} +   \frac{ i }{2 \sqrt{m E - \frac{3}{4} k^2}}    } .
\end{equation}
In addition to finding solutions of Eq.(\ref{TiLeq1D}) in $i L$ space for discrete momenta values, Eq.(\ref{TiLeq1D}) also allow us  to  analytic continue argument $k$ in $\mathcal{T}_{iL} (k; E ) $ into real continuous values that can be used to compute Dalitz plot {\it etc}. 

First of all, Eq.(\ref{TiLeq1D}) can be solved for $k \in \frac{2\pi n}{i L}, n \in \mathbb{Z}$ by matrix inversion, 
\begin{equation}
\mathcal{T}_{iL} (p; E ) = \sum_{k  } \left [ \mathcal{D}_{i L}^{-1} (E) \right ]_{p, k}    \tau^{(k)}_{iL} (E) , \label{TiLipsol1D}
\end{equation}
where  $(k,p) \in \frac{2\pi n}{i L}, n \in \mathbb{Z}$,  and $\mathcal{D}_{i L} (E)$ matrix is given by
\begin{equation}
 \left [ \mathcal{D}_{i L} (E) \right ]_{p, k} = \delta_{p, k} + 2 \tau_{iL}^{(k)} (E)    (i L ) \widetilde{G}_{iL} (k,p; E) .
\end{equation}
Next, plugging Eq.(\ref{TiLipsol1D}) into Eq.(\ref{TiLeq1D}),  therefore, the $\mathcal{T}_{iL} (k; E ) $ with real continuous $k$ argument is now obtained by
\begin{equation}
 \mathcal{T}_{iL} (k; E ) = \tau_{iL}^{(k)} (E) g_{i L} (k; E)  , \label{TiLsol1D}
\end{equation}
where
\begin{align}
&  g_{i L} (k; E) =  1 \nonumber \\
&-2    \sum_{ (p,p') = \frac{2\pi n}{i L}, n \in \mathbb{Z}}   (i L ) \widetilde{G}_{iL} (k,p; E)  \left [ \mathcal{D}_{i L}^{-1} (E) \right ]_{p, p'}    \tau^{(p')}_{iL} (E) . \label{giL3b}
\end{align}
The second term in $ g_{i L} $ function   describes the correction to isobar model from crossed channels due to rescattering effect.

\begin{figure}
\begin{center}
\includegraphics[width=0.88\textwidth]{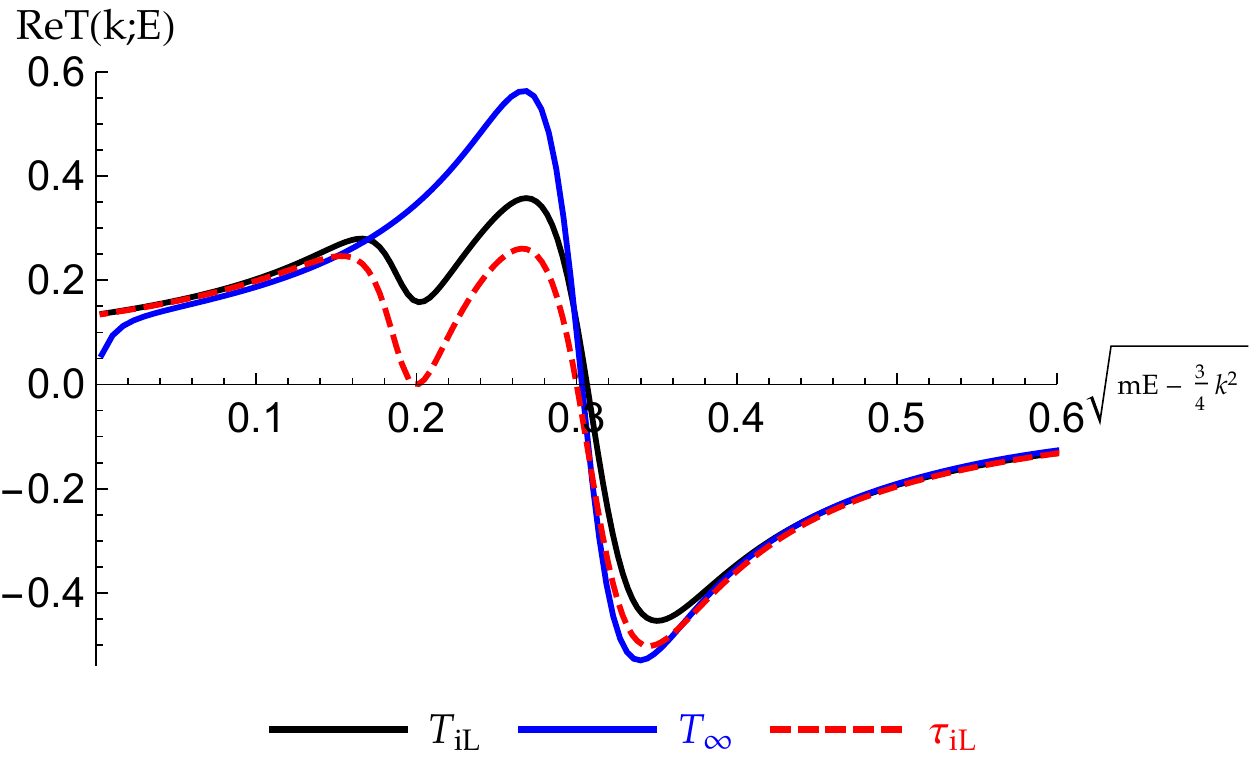}
\includegraphics[width=0.88\textwidth]{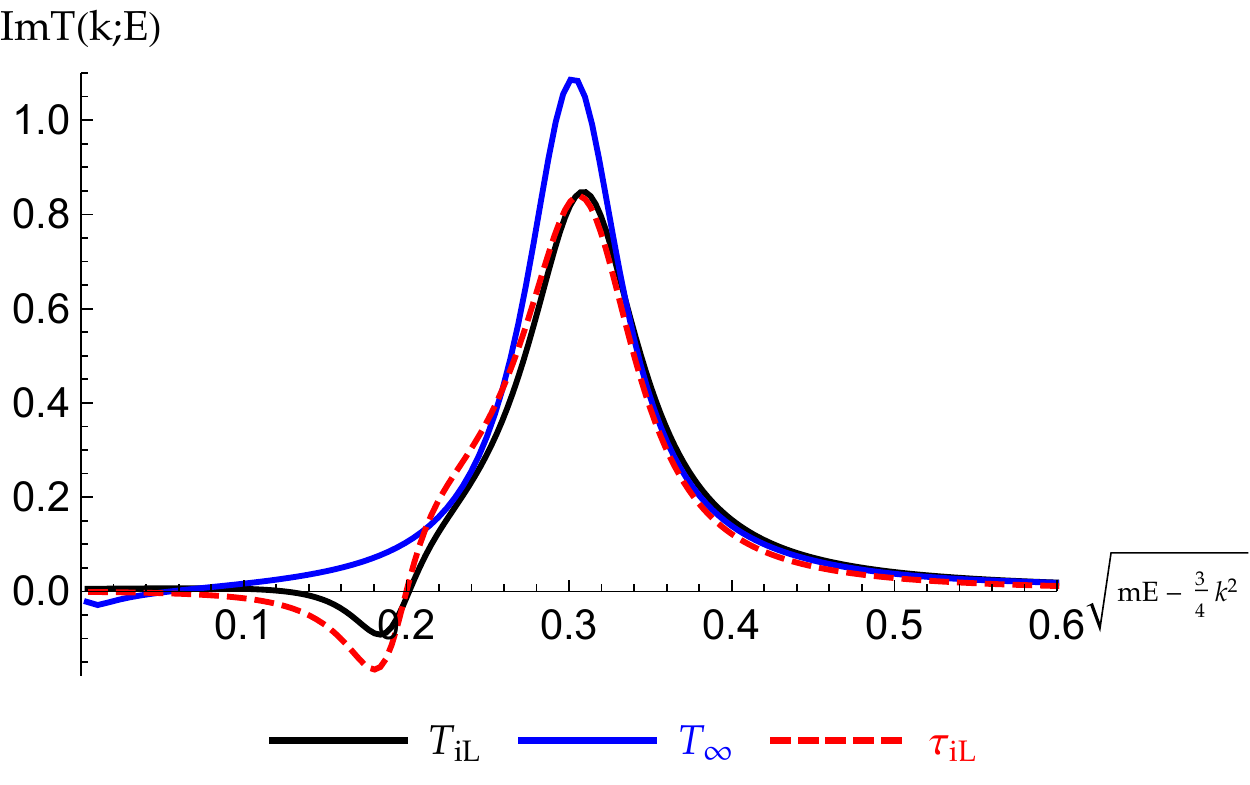}
\caption{ The comparison of    $ \mathcal{T}_{iL} (k; E )$ (solid black), $\tau^{(res, k)}_{iL} (E) $ (solid blue) and $ \mathcal{T}_{\infty} (k; E ) $  with fixed $m E =0.8$,    $L=10 $, $mV_0=0$, $g_\rho=0.04$ and $m_\rho=0.3$. }\label{T3bresfvinfvplot}
\end{center}
\end{figure}

Again, with a simple  resonance model of two-body amplitude by replacement $  m V_0  \rightarrow m V_0 + \frac{g_\rho}{m E- \frac{3}{4} k^2 - m  m_\rho  }$, and $\tau^{( k)} (E)  \rightarrow \tau^{(res, k)} (E) $ in both finite and infinite volumes, for instance
\begin{align}
&  \left [ \tau^{(res, k)}_{iL} (E) \right ]^{-1}=    -   \frac{1}{ m V_0 + \frac{g_\rho}{m E- \frac{3}{4} k^2 - m  m_\rho  } }   \nonumber \\
&- i  \frac{ \coth \frac{ \sqrt{m E - \frac{3}{4} k^2} - \frac{k}{2}}{2} L + \coth \frac{ \sqrt{m E - \frac{3}{4} k^2} + \frac{k}{2}}{2} L  }{4 \sqrt{m E - \frac{3}{4} k^2}}      ,
\end{align}
 thus the comparison of numerical solutions of $ \mathcal{T}_{iL} (k; E ) $ given by Eq.(\ref{TiLsol1D}) and  its counterpart in infinite volume, $\mathcal{T}_{\infty} (k; E )$, given by Eq.(\ref{Tinfveq1D}) is shown in Fig.~\ref{T3bresfvinfvplot}.   

 \begin{figure}
\begin{center}
\includegraphics[width=0.88\textwidth]{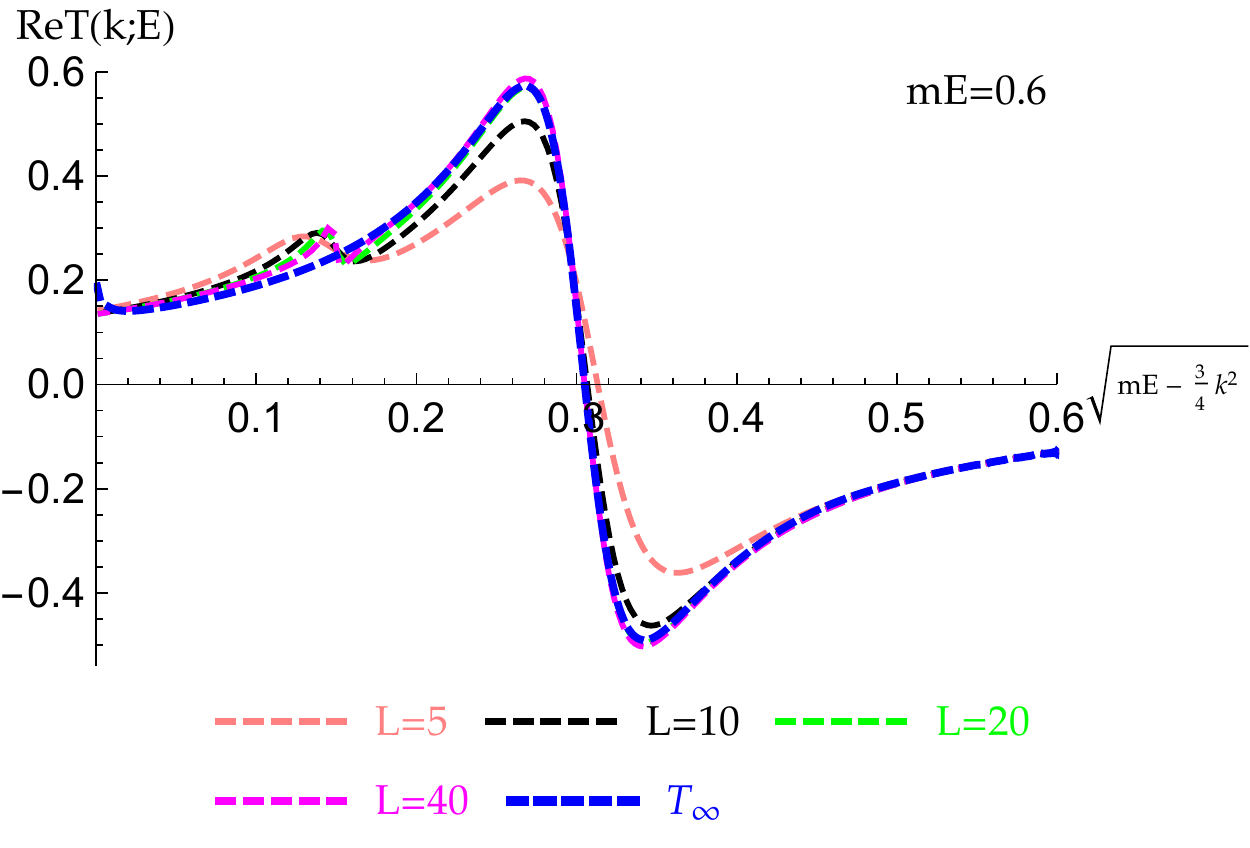}
\includegraphics[width=0.88\textwidth]{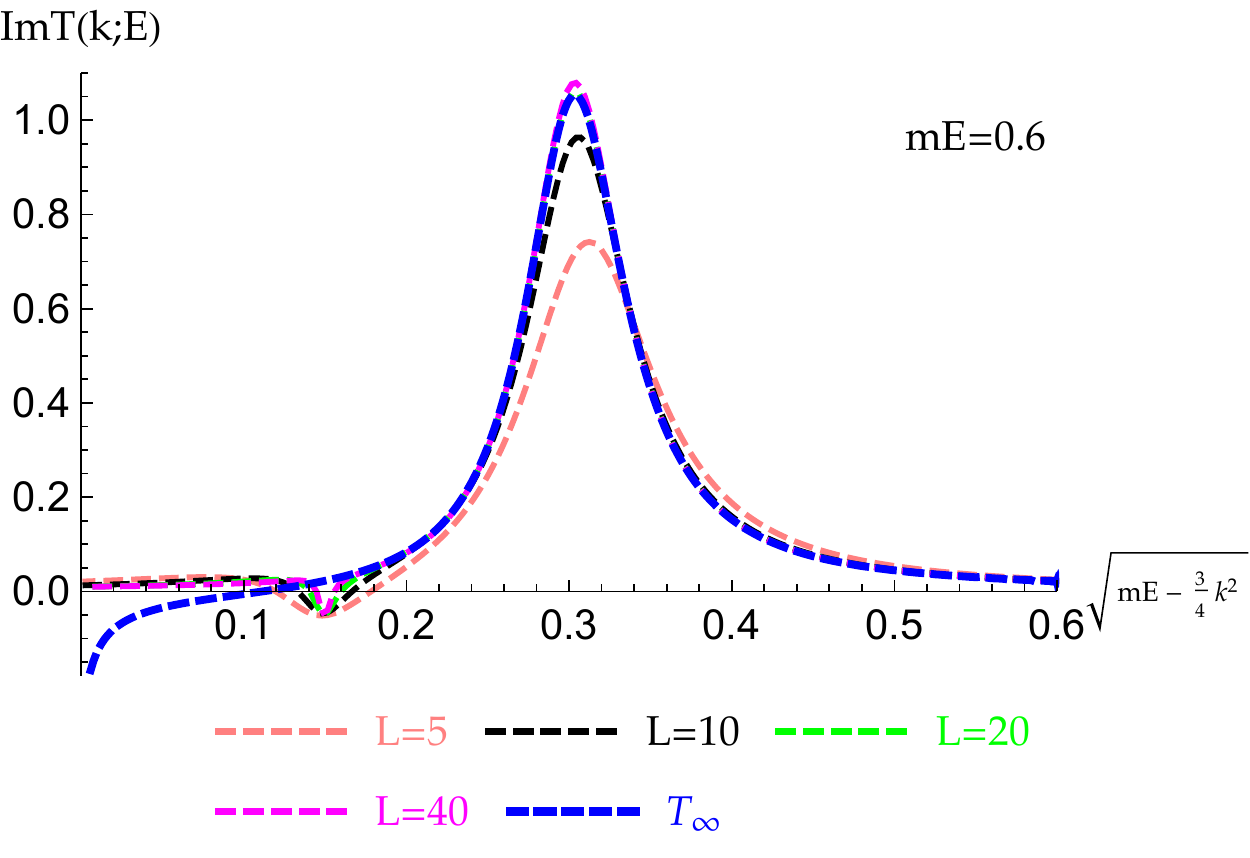}
\caption{ The comparison of    $ \mathcal{T}_{iL} (k; E )$ and $ \mathcal{T}_{\infty} (k; E ) $ (blue)   with  multiple $L$'s:  $L=5$ (pink), $10$ (black), $20$ (green),  $40$ (magenta). Model parameters are fixed with $m E =0.6$,     $mV_0=0$, $g_\rho=0.04$ and $m_\rho=0.3$. }\label{T3bfvLsE0dot6plot}
\end{center}
\end{figure}
  
 \begin{figure}
\begin{center}
\includegraphics[width=0.88\textwidth]{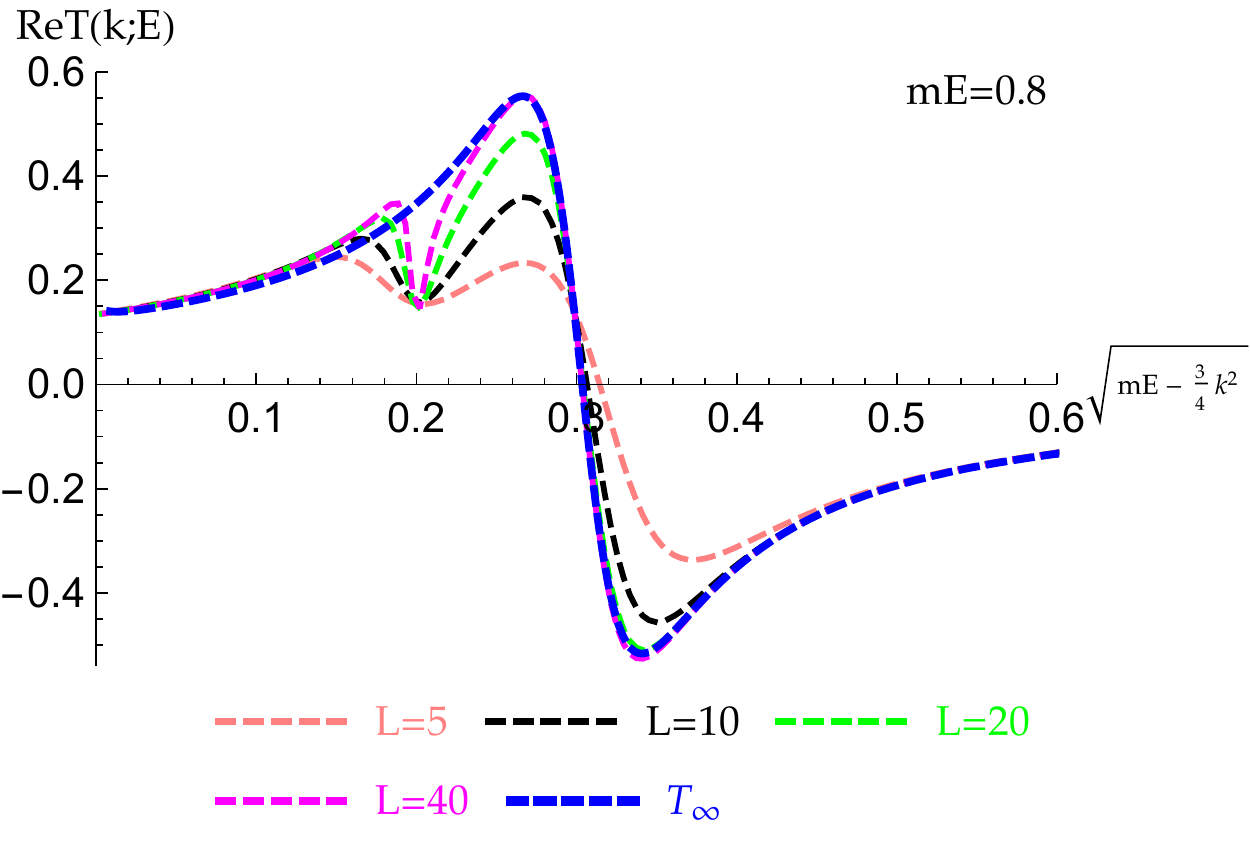}
\includegraphics[width=0.88\textwidth]{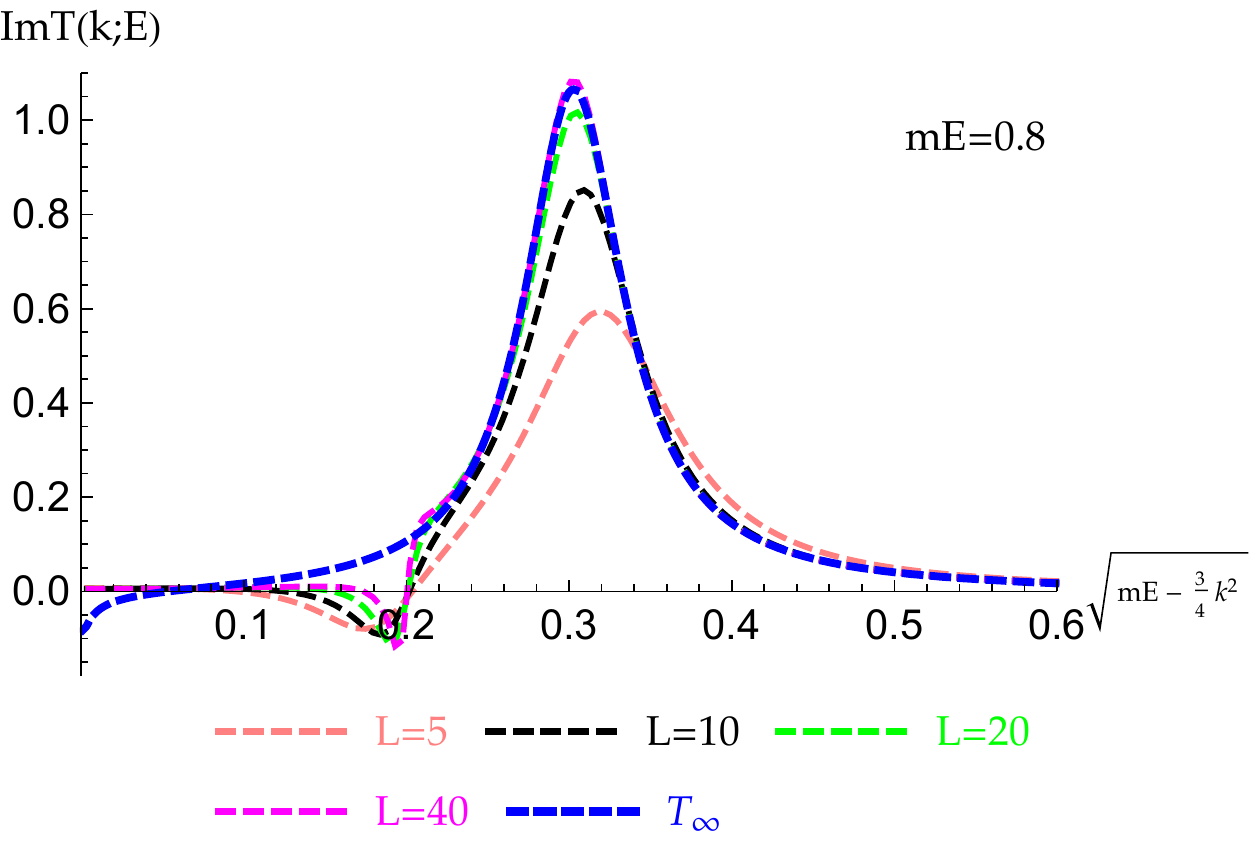}
\caption{ The comparison of    $ \mathcal{T}_{iL} (k; E )$ and $ \mathcal{T}_{\infty} (k; E ) $ (blue)   with a  fixed   $m E =0.8$ and  multiple $L$'s:  $L=5$ (pink), $10$ (black), $20$ (green),  $40$ (magenta). }\label{T3bfvLsplot}
\end{center}
\end{figure}

  \begin{figure}
\begin{center}
\includegraphics[width=0.88\textwidth]{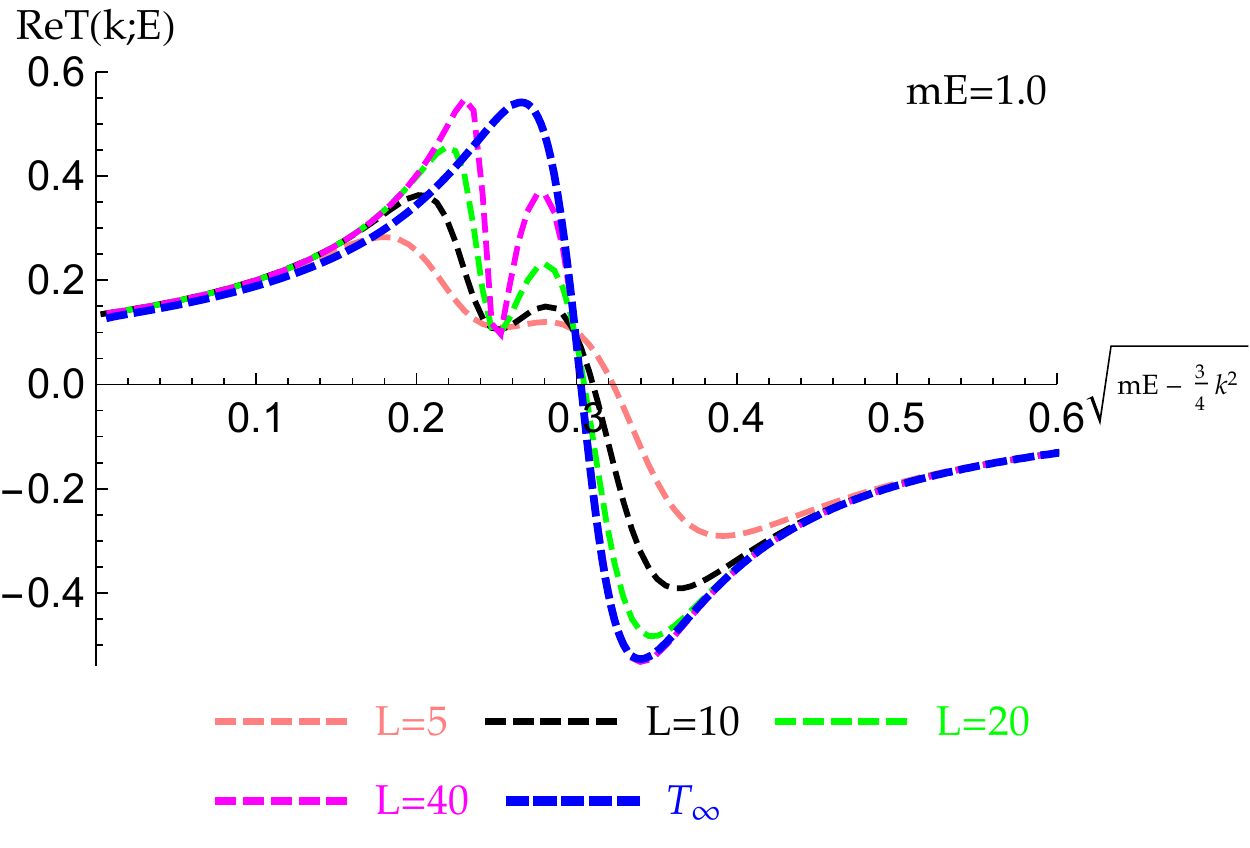}
\includegraphics[width=0.88\textwidth]{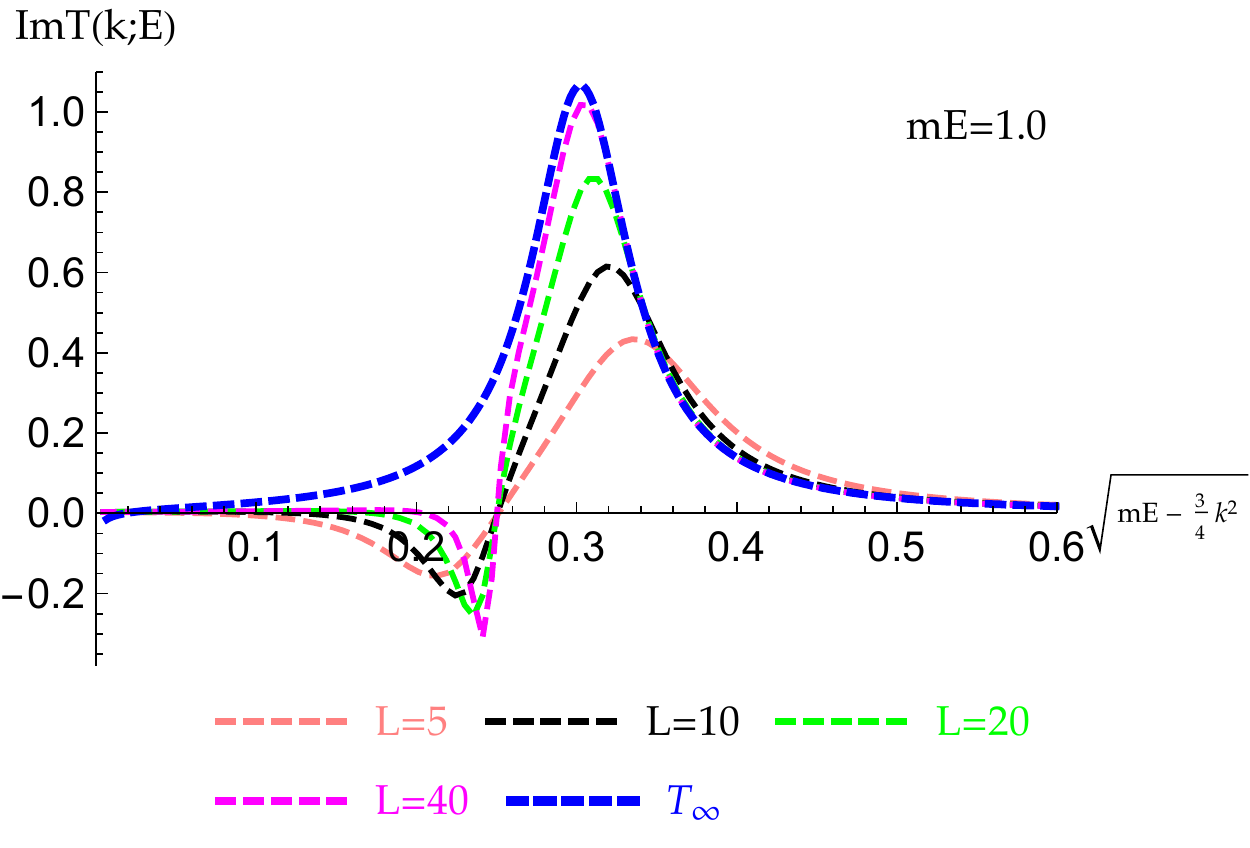}
\caption{ The comparison of    $ \mathcal{T}_{iL} (k; E )$ and $ \mathcal{T}_{\infty} (k; E ) $ (blue)   with    a  fixed   $m E =0.8$ and  multiple $L$'s:  $L=5$ (pink), $10$ (black), $20$ (green),  $40$ (magenta). }\label{T3bfvLsE1dot0plot}
\end{center}
\end{figure}

We remark that the cusp effect in $ \mathcal{T}_{iL} (k; E ) $ for real $k$  is pure finite volume artifact, also see Fig.~\ref{T3bfvLsE0dot6plot},  \ref{T3bfvLsplot}    and  \ref{T3bfvLsE1dot0plot}   for the plot of $ \mathcal{T}_{iL} (k; E ) $ with multiple $L$'s and $mE$'s. The finite volume cusp  originates from   analytic continuation of  the finite volume three-body Green's function with real arguments,
\begin{align}
 \sum_{ p = \frac{2\pi n}{i L}, n \in \mathbb{Z}} &   (i L ) \widetilde{G}_{iL} (k,p; E)  = \frac{1}{i L} \frac{1}{mE - k^2} \nonumber \\
& + \frac{1}{iL}   \sum_{ p = \frac{2\pi n}{i L},n \neq 0} \frac{1}{ m E - p^2 - p k - k^2} .
\end{align}
Now, we can clearly see the pole term in finite volume Green's function, $\frac{1}{mE -k^2}$, which  appears to cause trouble for the real  $k$ values near $\pm \sqrt{mE}$.  In fact, this singular term shows up in   $g_{iL}(k;E)$ defined in Eq.(\ref{giL3b}) as pole singularity,    however it also shows up in 
$$ \tau^{(k)}_{iL} (E) = -  \frac{1}{ \frac{1}{mV_0} -  \sum_{ p = \frac{2\pi n}{i L}}^{n \in \mathbb{Z}}    (i L ) \widetilde{G}_{iL} (k,p; E)   }  \propto (mE-k^2) $$ as zero, see  red dashed curves in Fig.~\ref{T3bresfvinfvplot}. Therefore, finite volume three-body amplitude  $\mathcal{T}_{iL} (k; E ) = \tau_{iL}^{(k)} (E) g_{i L} (k; E) $ is free of singularity in the end.   Although the singularity is canceled out, the  finite volume effect  still   shows up as a cusp which is absent in infinite volume amplitudes. We also notice that the   finite volume cusp is located at $k \sim \pm \sqrt{mE}$ while the location of two-body resonance is   around  $ mE -\frac{3}{4} k^2 \sim m m_{\rho}$. Hence the cusp may start interfering with the shape of resonance   when $ E \sim 4  m_\rho $, see Fig.~\ref{T3bfvLsE0dot6plot},  \ref{T3bfvLsplot}    and  \ref{T3bfvLsE1dot0plot} as  an example.

\section{Summary and discussion}\label{summary} 
  In summary,  we explore and experiment an alternative approach of studying resonance properties in finite volume. As the consequence of global  symmetry of Green's function in complex spatial plane, by analytic continuation of $L$ into $i L$, the oscillating behavior of finite volume Green's function can be mapped into infinite volume Green's function with corrections of exponentially decaying finite volume effect.  Using finite volume size $L$ as a tuning knob, thus  the finite volume scattering amplitude may behave similarly to infinite volume amplitude in $i L$ space. A cusp in three-body finite volume amplitude due to finite volume effect is also observed, it may start interfering with and distorting the shape of resonance  while total energy $E$ is in certain range.
  Nevertheless the resonance peak is still clearly visible even in a small box. Hence, the resonance properties may be computed directly from finite volume dynamical equations.

  The  analytical continuation technique  presented in the paper may be useful in visualizing  resonances  from finite volume dynamical equations, in processes such as $\eta , \eta' \rightarrow 3\pi$. In order to describe three-particle resonance, three-body short-range interaction must be included as well. a simple model that  may be used to describe three-particle resonance  is sketched in Appendix \ref{3particleresonance}.

\begin{acknowledgements}
 We   acknowledge support from the Department of Physics and Engineering, California State University, Bakersfield, CA.   This research was supported in part by the National Science Foundation under Grant No. NSF PHY-1748958.   P.G.   acknowledges GPU computing resources (http://complab.cs.csubak.edu) from  the Department of Computer and Electrical Engineering and Computer Science at California State University-Bakersfield   made available for conducting the research reported in this work.  B.L. acknowledges support     by the National Science Foundation of China under Grant Nos.11775148 and 11735003.
 \end{acknowledgements}

\appendix

\section{Multiparticle wave function in finite volume}\label{wavanlycont}
 
The total multiparticle wave function is related to the total finite volume amplitude by
\begin{equation}
\Phi_L  =  - G_L  (E) T_L^{tot} (E).
\end{equation}
The total amplitude $T_L^{tot} (E)$ is usually given by the sum of multiple terms,  for instance, in three-body interaction with only pair-wise interaction, thus
\begin{equation}
T_L^{tot} (E) = \sum_{\gamma=1}^3 T_L^{(\alpha \beta)} (E), \ \ \ \  \alpha \neq \beta \neq \gamma,
\end{equation}
where $ T_L^{(\alpha \beta )} (E)$ satisfies coupled homogeneous equations, such as
\begin{equation}
T_L^{(12)}  (E)= \frac{1}{\frac{1}{V_{12}}- G_L (E)} G_L(E)  \left [ T_L^{(23)}   (E)+ T_L^{(31)}  (E) \right ] .
\end{equation}
For three identical particles, thus coupled homogeneous equations are reduced to Eq.(\ref{TLeq})  and    $ T_L^{(\alpha \beta)} (E)  = T_L (E)$.  Therefore, finding eigensolutions of finite volume Faddeev amplitude becomes a key step for computing wave function.

In general, the dynamical equations of finite volume Faddeev amplitudes may be casted as a  matrix equation in a linear form,
\begin{equation}
T_i (E) =\sum_j  K_{i,j} (E) T_j (E), \label{Tgeneq}
\end{equation}
where  the vector $T(E) $ stands for the energy dependent amplitude   that also depends on the discrete momenta,  the matrix $K (E) $ represents the energy dependent kernel function. In finite volume,  due to the periodic boundary condition, the Eq.(\ref{Tgeneq}) can be satisfied only for some discrete energies, $\{E_1,  \cdots, E_n , \cdots \}$.  Eq.(\ref{Tgeneq})  has non-trivial solutions only if
\begin{equation}
\det \left [ I - K (E) \right ] =0, \ \ \ \ E \in \{ E_n \}.
\end{equation}
In order to find eigenvector solution of Eq.(\ref{Tgeneq}), let's consider the subtracted equation of Eq.(\ref{Tgeneq}), 
\begin{equation}
T_i (E) = T_{i_0}  (E)  + \sum_j  \left [ K_{i,j} (E) -  K_{i_0,j}  (E) \right ] T_j (E), \label{subTgeneq}
\end{equation}
where $T_{i_0}  (E)$  stands for the $i_0$-th   element of $T(E)$ vector we choice for the subtraction.  $T_{i_0}  (E)$ may also be used as normalization factor and is a constant value for fixed $E$. Hence, the solution of subtracted Eq.(\ref{subTgeneq}) is obtained  by matrix inversion,
\begin{equation}
T(E) =  \frac{1}{ I - K (E) + K_0 (E) } T_0 (E), \label{Tgensol}
\end{equation}  
where $T_0 (E) = T_{i_0}  (E)$ vector is a constant for a fixed $E$,  and  $ [ K_0 ]_{i , j} (E) = K_{i_0,j}  (E) $. 
The expression in Eq.(\ref{Tgensol}) is indeed the eigenvector solution of Eq.(\ref{Tgeneq}) when $E \in \{ E_n \}$, since    $T(E_n) = K (E_n) T(E_n)$ and so is $T_0(E_n) = K_0 (E_n) T(E_n)$.
 The subtracted homogeneous dynamical equations hence can be used to find eigenvector solutions of finite volume Faddeev type equations.

Next, we will just use  a simple   example to illustrate above described approach. Let's consider three non-relativistic identical bosons interacting with contact interactions in 1D. The wave function is given in terms of  finite volume Faddeev amplitude by
\begin{align}
\phi_L &  (r_{13} , r_{23})  =  -  \sum_{p_1, p_2} e^{i p_1 r_{13}} e^{i p_2 r_{23}} \widetilde{G}_L (p_1,p_2; E)    \nonumber \\
& \times \left [ T_L(p_1;E) +T_L(p_2;E) + T_L(p_3;E) \right ] , \label{wav3b1D}
\end{align}
where $(p_1, p_2) \in \frac{2\pi}{L} n, n \in \mathbb{Z}$,  and $p_3 =- p_1 - p_2$. $r_{ij } $ are relative coordinates between i-th and j-th particles. 
The $T_L (E)$ satisfies integral equation,
\begin{equation}
T_L (k; E) = -2 \tau_L^{(k)} (E)   \sum_{p}  L \widetilde{G}_L (k,p; E)  T_L (p;E), \label{TLeq1D}
\end{equation}
where $(k, p) \in \frac{2\pi}{L} n, n \in \mathbb{Z}$. 
The eigenenergies are obtained by   quantization condition
\begin{equation}
\det \left [ \delta_{k, p} - K(k, p; E) \right ] =0,
\end{equation}
where kernel function is
\begin{equation}
K(k, p; E) =- 2 \tau_L^{(k)} (E)   L \widetilde{G}_L (k,p; E) .
\end{equation}
Once eigenenergies are determined, the eigenvector can be found by matrix inversion of subtracted Eq.(\ref{TLeq1D}),
\begin{equation}
  \sum_p \left [ \delta_{k ,p }  - K(k, p; E) + K(k_0, p; E) \right ] T_L (p;E) =  T_L (k_0;E),
\end{equation}
where $k_0$ is  the subtraction point and can be chosen arbitrarily, and $T_L (k_0)$ may be treated as normalization factor.

\subsection{Multiparticle energy spectrum of a resonance model}\label{resspec}

\begin{figure}
\begin{center}
\includegraphics[width=0.98\textwidth]{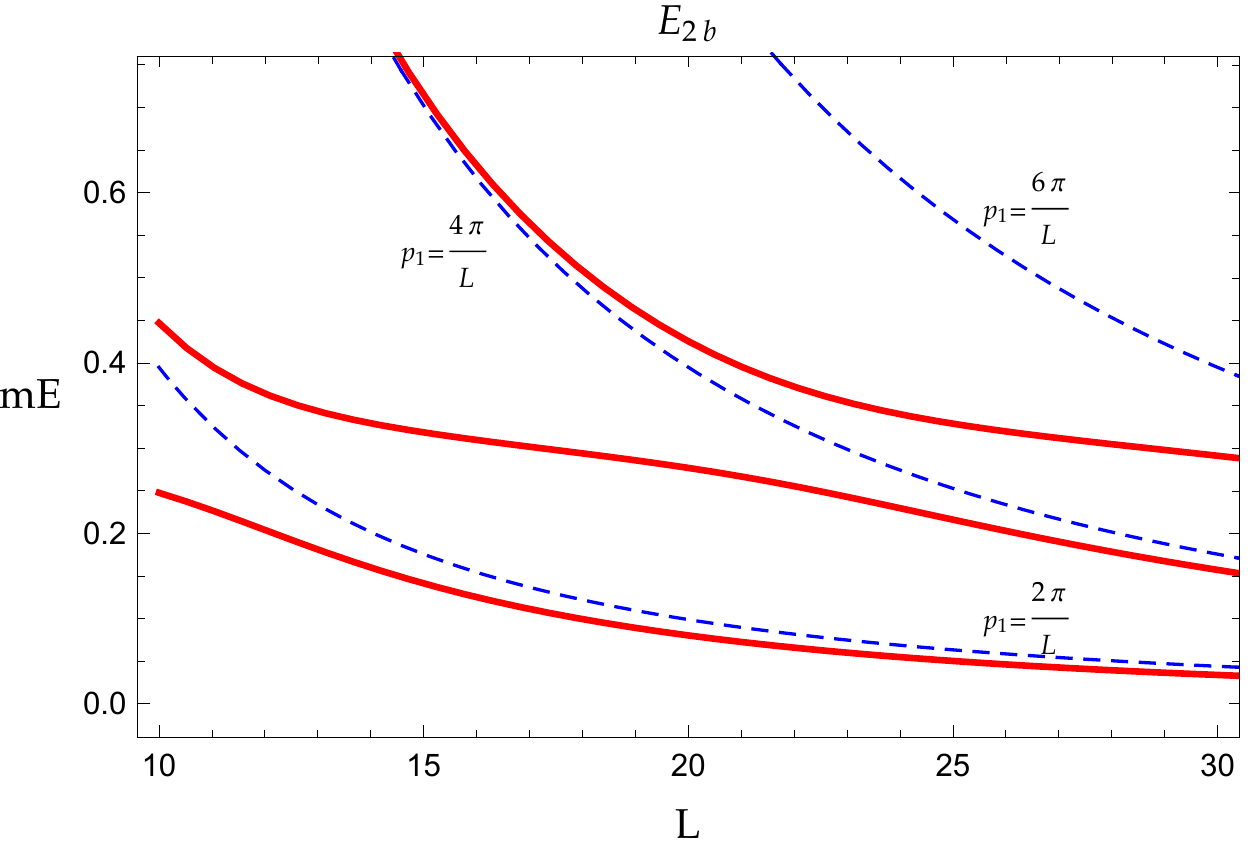}
\caption{  The CM frame two particles energy spectrum (solid red) for a resonance model with  parameters: $mV_0=0$, $g_\rho=0.04$ and $m m_\rho=0.3 $. The free particles energy spectrum (dashed blue) are also plotted as reference with $m  E_{2b}^{(free)} = (\frac{2\pi n}{L} )^2$, $n \in \mathbb{Z}$. }\label{E2bplot}
\end{center}
\end{figure}

\begin{figure}
\begin{center}
\includegraphics[width=0.98\textwidth]{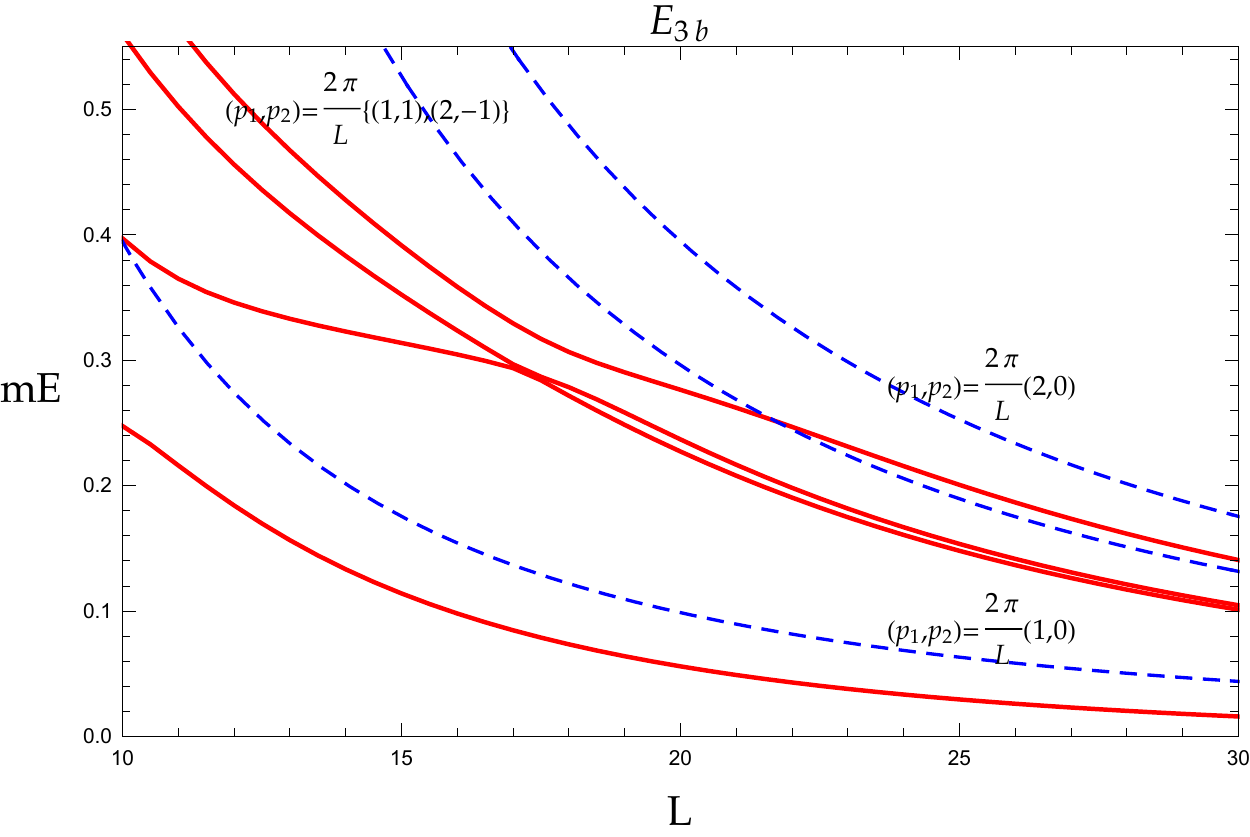}
\caption{ The CM frame three particles energy spectrum (solid red) for a resonance model with  parameters: $mV_0=0$, $g_\rho=0.04$ and $m m_\rho=0.3$. The free particles energy spectrum (dashed blue) are also plotted as reference with $m E_{3b}^{(free)} =  p_1^2 + p_1 p_2 + p_2^2  $, $ (p_1,p_2) \in \frac{2\pi n}{L},  n \in \mathbb{Z}$. }\label{E3bplot}
\end{center}
\end{figure}

To make it more interesting, let's   propose a resonance model with following replacement in contact interaction,
\begin{equation}
m V_0  \rightarrow m V_0   +  \frac{g_\rho}{m E -\frac{3}{4} k^2 -  m m_\rho }, \label{vrescoup}
\end{equation}
 where $(g_\rho, m_\rho)$ stand for the coupling constant and mass of resonance respectively.  Therefore, the two- and three-particle energy spectrum in CM frame are determined  by two-body quantization condition
 \begin{equation}
     \frac{1}{m V_0 +  \frac{g_\rho}{m ( E -    m_\rho )} } -  \frac{ \cot \frac{ \sqrt{m E  }}{2} L }{2 \sqrt{m E  }}  =0  
  \end{equation}
  and three-body quantization condition
  \begin{equation}
\det \left [ \delta_{k, p} +  \frac{1}{L}   \frac{2 \tau_L^{(res,k)} (E) }{mE -k^2 - k p - p^2}   \right ] =0
\end{equation}
respectively, where 
 \begin{align}
&  \left [ \tau^{(res, k)}_{L} (E) \right ]^{-1}=    -   \frac{1}{ m V_0 + \frac{g_\rho}{m E- \frac{3}{4} k^2 - m  m_\rho  } }   \nonumber \\
&- i  \frac{ \coth \frac{ \sqrt{m E - \frac{3}{4} k^2} - \frac{k}{2}}{2} L + \coth \frac{ \sqrt{m E - \frac{3}{4} k^2} + \frac{k}{2}}{2} L  }{4 \sqrt{m E - \frac{3}{4} k^2}}      .
\end{align} 
The two- and three-particle energy spectra are shown in Fig.~\ref{E2bplot} and Fig.~\ref{E3bplot} respectively. The resonance shows up   at $m E_{2b} = 0.3$ location and flatten up the curves of two-body energy spectrum as the function of $L$ near $m E_{2b} =0.3$. Similar pattern is observed in three-body energy spectrum, however, the situation in three-body sector is slightly more sophisticated. The energy spectrum of free particles shows the degeneracy for some levels, such as    blue curve in  the middle has double degeneracy for $(p_1,p_2) = \frac{2\pi}{L} (1,1)$ and $(p_1,p_2) = \frac{2\pi}{L} (2,-1)$, hence, three blue free three-body energy spectrum curves in fact represent four energy levels. The degeneracy in second and third levels in the middle is   removed by interaction, see the splitting in two red curves in the middle in  Fig.~\ref{E3bplot}.

  \begin{figure}
\begin{center}
\includegraphics[width=0.98\textwidth]{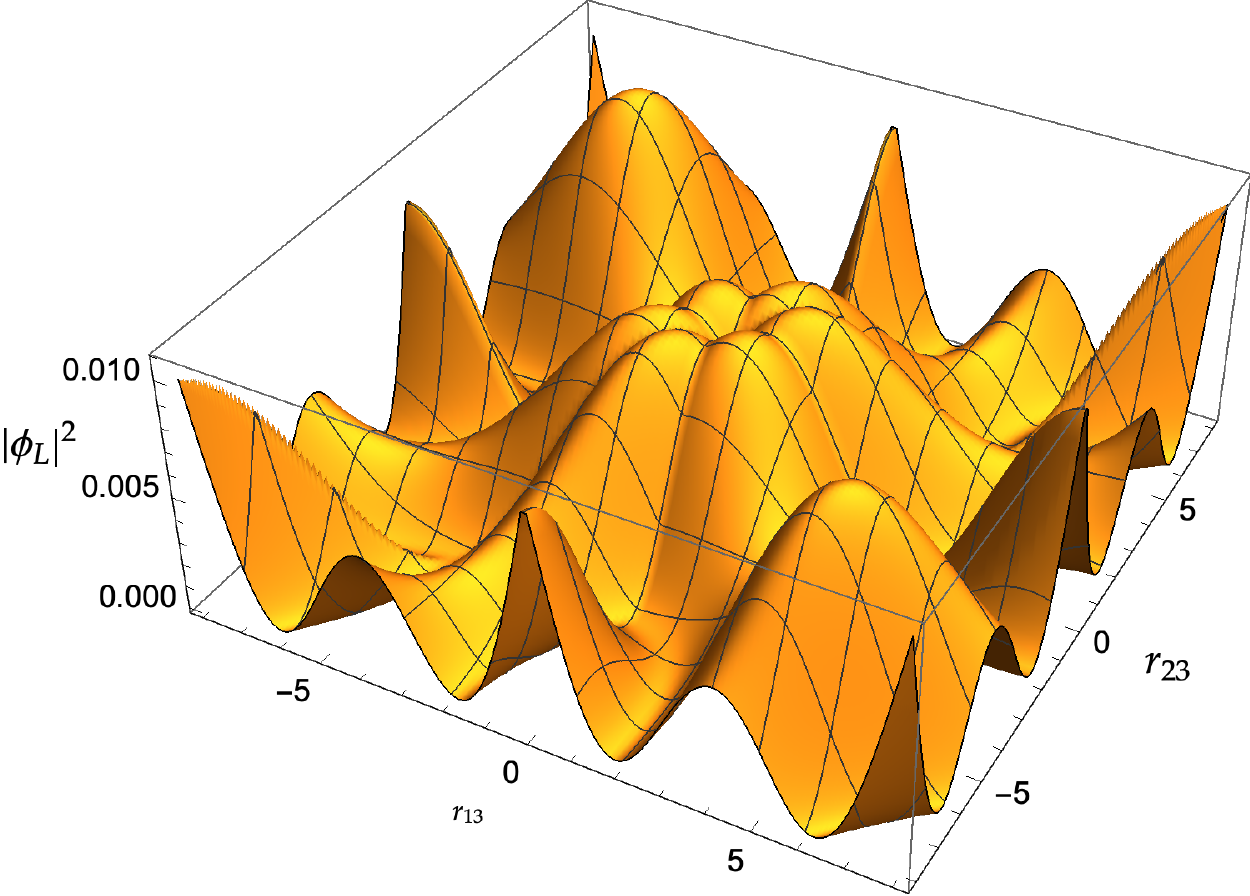}
\caption{  Three-body wave function $|\phi_{L} (r_{13}, r_{23})|^2$ for a resonance model with  parameters: $mV_0=0$, $g_\rho=0.04$ and $m m_\rho=0.3$, and  $L=16$ and   $m  E_{3b}  = 0.3046 \sim m m_\rho$. }\label{wavplot}
\end{center}
\end{figure}

\subsection{Three particles wave function of  a resonance model}
With the resonance model proposed in Sec.\ref{resspec}, we can thus apply the technique described  previously to compute three particles wave function. The three-body finite volume $T_L$ amplitude is solved by matrix inversion and is  given by 
\begin{equation}
   T_L (p;E) =\sum_k \left [  D^{-1} (k_0,E)\right ]_{p,k}  T_L (k_0;E),
\end{equation}
where $k_0$ is arbitrary subtraction point, and $D$ matrix is given by
\begin{equation}
D_{k, p}(k_0;E) =\delta_{k ,p }  - K(k, p; E) + K(k_0, p; E).
\end{equation}
Using $T_L (p;E) $ as input, thus the three-body wave function can be computed by Eq.(\ref{wav3b1D}), see  Fig.~\ref{wavplot} as a example with $mE_{3b} $ near resonance mass.

\section{Three-particle resonance model}\label{3particleresonance}
 In order to describe three-particle resonance decay or scattering process, the three-body short-range interaction must be included. In this section, we present  a simple three-body contact interaction model  that may be useful to describe process such as $\eta \rightarrow 3 \pi$. When three-body short-range interaction is included, all the discussion in presented in Sec.\ref{3bFaddeev} must be extended by including  a  three-body amplitude $T^{(3b)}$. Thus Eq.(\ref{Topoffsheeq}) is replaced by  coupled equations
\begin{equation}
\mathcal{T}_L (E)  =\tau_L(E)  -  \tau_L(E)   G_L (E) \left [ 2 \mathcal{T}_L (E)  +   \mathcal{T}^{(3b)}_L (E)  \right ],   \label{T3bFaddeeveq}
\end{equation}
and
\begin{equation}
  \mathcal{T}^{(3b)}_L (E)    =\tau^{(3b)}_L(E)  -  \tau^{(3b)}_L(E)   G_L (E) 3 \mathcal{T}_L (E)   ,\end{equation} 
where  $ \tau^{(3b)}_L$ is related to three-body interaction $V_{123}$ by
\begin{align}
  \tau^{(3b)}_L(E) =  - \frac{1}{\frac{1}{V_{123}} - G_L (E)} .
\end{align}
With the same convention, $\tau_L$ is related to two-body pair-wise interaction $V$ by
\begin{align}
\tau_L(E) =  - \frac{1}{\frac{1}{V} - G_L (E)} . 
\end{align}
Eliminating $ \mathcal{T}^{(3b)}_L $ in Eq.(\ref{T3bFaddeeveq}), we thus find
\begin{align}
& \mathcal{T}_L (E)  =\tau_L(E)  \left[  1-G_L (E)\tau^{(3b)}_L(E)   \right ] \nonumber \\
& -  \tau_L(E)   G_L (E) \left [ 2  - 3  \tau^{(3b)}_L(E)   G_L (E)    \right ]  \mathcal{T}_L (E)  .  \label{T3bFaddeevredeq}
\end{align}
The quantization condition Eq.(\ref{3bqcpairwise}) is thus replaced by
\begin{equation}
\det \left [ I+ \tau_L(E)  G_L  (E)  \left ( 2  - 3  \tau^{(3b)}_L(E)   G_L (E)    \right )  \right ] =0,
\end{equation}
where  three-body term $  \tau^{(3b)}_L(E)  $  may be modeled to describe the impact of three-body resonance to the spectrum of three-body system.

In momentum basis with a short-range three-body interaction, Eq.(\ref{T3bFaddeevredeq}) is thus given by
\begin{align}
& \mathcal{T}_L (k; E ) = \tau_L^{(k)} (E)  \left [ 1-  \bigg ( \sum_{p'}L \widetilde{G}_L (k,p'; E)  \bigg  )  \tau^{(3b)}_L (E)   \right ]    \nonumber \\
& -  \tau_L^{(k)} (E)   \sum_{p  } \left [ 2  L \widetilde{G}_L (k,p; E)  -3    \bigg ( \sum_{p'}L \widetilde{G}_L (k,p'; E)  \bigg  )  \right. \nonumber \\
& \quad \quad  \quad \quad  \quad \quad \times \left.  \tau^{(3b)}_L (E)   \bigg ( \sum_{p''}L \widetilde{G}_L (p'',p; E)  \bigg  )  \right ]  \mathcal{T}_L (p;E ),
\end{align}
where
\begin{align}
  \tau^{(3b)}_L(E) =  - \frac{1}{\frac{1}{m V_{123}} -  \sum_{k,p}  \widetilde{G}_L (k,p; E) } .
\end{align}
The three-particle resonance may be modeled by replacing   $V_{123}$ by
$$m V_{123} + \frac{g_{123}}{m (E- m_{123})} $$ where $g_{123}$ and $m_{123}$ represent the coupling strength and mass of three-body resonance.

\bibliography{ALL-REF.bib}

\end{document}